\documentclass[review,onefignum,onetabnum]{siamart171218}

\usepackage{multirow}
\usepackage{lipsum}
\usepackage{amsfonts}
\usepackage{graphicx}
\usepackage{epstopdf}
\usepackage{algorithmic}
\ifpdf
  \DeclareGraphicsExtensions{.eps,.pdf,.png,.jpg}
\else
  \DeclareGraphicsExtensions{.eps}
\fi

\newsiamremark{remark}{Remark}
\newsiamremark{hypothesis}{Hypothesis}
\crefname{hypothesis}{Hypothesis}{Hypotheses}
\newsiamthm{claim}{Claim}

\title{Robust iterative method for symmetric quantum signal processing in all parameter regimes}

\author{Yulong Dong \thanks{Department of Mathematics, University of California,
Berkeley, CA 94720, USA. Email: \texttt{dongyl@berkeley.edu}} \and 
Lin Lin \thanks{Department of Mathematics and Challenge Institute for Quantum Computation, University of California, Berkeley, and Applied Mathematics and Computational Research Division, Lawrence Berkeley
National Laboratory, Berkeley, CA 94720, USA. Email:
\texttt{linlin@math.berkeley.edu}}\and
Hongkang Ni \thanks{Institute for Computational and Mathematical Engineering (ICME), Stanford University, Stanford, CA 94305,
USA. Email: \texttt{hongkang@stanford.edu}}\and
Jiasu Wang \thanks{Department of Mathematics, University of California,
Berkeley, CA 94720, USA. Email: \texttt{jiasu@berkeley.edu}
}
}

\usepackage{amsopn}

\makeatletter
\newcommand*{\addFileDependency}[1]{
  \typeout{(#1)}
  \@addtofilelist{#1}
  \IfFileExists{#1}{}{\typeout{No file #1.}}
}
\makeatother

\usepackage{algorithm}
\usepackage{algorithmic}
\usepackage{subcaption}
\usepackage{braket}
\usepackage{amsmath}
\usepackage{amssymb}
\usepackage{tikz}
\usetikzlibrary{decorations.pathreplacing}
\usepackage{listings}

\newcommand{\bvec}[1]{\mathbf{#1}}

\newcommand{\vg}{\bvec{g}}

\newcommand{\vv}{\bvec{v}}

\renewcommand{\Re}{\mathrm{Re}}
\renewcommand{\Im}{\mathrm{Im}}
\newcommand{\I}{\mathrm{i}}

\newcommand{\mc}[1]{\mathcal{#1}}

\newcommand{\wt}[1]{\widetilde{#1}}

\newcommand{\abs}[1]{\left\lvert#1\right\rvert}
\newcommand{\norm}[1]{\left\lVert#1\right\rVert}
\newcommand{\argmin}{\mathop{\mathrm{argmin}}}

\newcommand{\ud}{\,\mathrm{d}}

\newcommand{\NN}{\mathbb{N}}
\newcommand{\RR}{\mathbb{R}}
\newcommand{\CC}{\mathbb{C}}

\newcommand{\FF}{\mathbb{F}}

\newcommand{\Or}{\mathcal{O}}

\begin{document}
\nolinenumbers
\maketitle

\begin{abstract}
 This paper addresses the problem of solving nonlinear systems in the context of symmetric quantum signal processing (QSP), a powerful technique for implementing matrix functions on quantum computers. Symmetric QSP focuses on representing target polynomials as products of matrices in SU(2) that possess symmetry properties. We present a novel Newton's method tailored for efficiently solving the nonlinear system involved in determining the phase factors within the symmetric QSP framework. Our method demonstrates rapid and robust convergence in all parameter regimes, including the challenging scenario with ill-conditioned Jacobian matrices, using standard double precision arithmetic operations. For instance, solving symmetric QSP for a highly oscillatory target function $\alpha \cos(1000 x)$ (polynomial degree $\approx 1433$) takes $6$ iterations to converge to machine precision when $\alpha=0.9$, and the number of iterations only increases to $18$ iterations when $\alpha=1-10^{-9}$ with a highly ill-conditioned Jacobian matrix.
Leveraging the matrix product states the structure of symmetric QSP, the computation of the Jacobian matrix incurs a computational cost comparable to a single function evaluation. Moreover, we introduce a reformulation of symmetric QSP using real-number arithmetics, further enhancing the method's efficiency. Extensive numerical tests validate the effectiveness and robustness of our approach, which has been implemented in the QSPPACK software package. 

\end{abstract}

\section{Introduction}

Many scientific computing problems can be viewed as implementing matrix functions $A \mapsto f(A)$. For simplicity we can assume that $A$ is a Hermitian matrix with eigenvalues in $[-1,1]$, and that $f:\RR\to\RR$ is a real polynomial. Quantum signal processing (QSP)~\cite{LowChuang2017,LowChuang2019,GilyenSuLowEtAl2019,MartynRossiTanEtAl2021} provides a systematic approach and a compact quantum circuit for implementing a broad class of matrix functions on quantum computers. This leads to efficient algorithms for various quantum applications, including linear system solving, Hamiltonian system simulation, ground-state energy estimation, and quantum benchmarking~\cite{LowChuang2017,GilyenSuLowEtAl2019,LinTong2020,MartynRossiTanEtAl2021,DongLinTong2022,DongLin2021,DongWhaleyLin2021,McArdleGilyenBerta2022,DongGrossNiu2022}. Recently, the construction of QSP has been studied and generalized using advanced theoretical tools~\cite{RossiChuang2022,RossiChuang2023,RossiBastidasMunroEtAl2023}.

Since any continuous function can be approximated using polynomials, the key idea behind QSP is to represent a target polynomial as a product of matrices in the special unitary group $\mathrm{SU}(2)$, parameterized by a set of phase factors denoted as $\Phi$. However, due to the constraints of $\mathrm{SU}(2)$, the target polynomial must satisfy specific conditions. 
\begin{definition}[Target polynomial of QSP]
A polynomial $f\in\RR[x]$ is called a \emph{target polynomial} of quantum signal processing if it satisfies (1) $\deg(f)=d$, (2) the  parity of $f$ is $(d \bmod 2)$, (3) $\norm{f}_{\infty}:=\max_{x\in[-1,1]} \abs{f(x)}\le 1$. Furthermore, $f$ is called \emph{fully-coherent} if $\norm{f}_{\infty}=1$.
\label{def:target_poly}
\end{definition}

The mapping from the target polynomial of degree $d$ (described by its Chebyshev coefficients denoted by $c\in\RR^{d+1}$) to phase factors $\Psi\in\RR^{d+1}$ can be abstractly written as
\begin{equation}\label{eq:obj}
    F(\Psi)=c.
\end{equation}
The mapping $F$ is highly nonlinear and is not one-to-one. For a given $c$, and our goal is to find \textit{a} solution to the nonlinear system \eqref{eq:obj}.

QSP in the fully-coherent regime (or near fully-coherent regime,  where $\norm{f}_{\infty}=1-\delta$ for a small $\delta>0$) finds applications in quantum algorithms for Hamiltonian simulation~\cite{MartynLiuChinEtAl2023} and time-marching based simulation of non-Hermitian dynamics~\cite{FangLinTong2023}. 
As will be discussed below, this problem is particularly challenging in the fully-coherent regime where the Jacobian matrix of $F$ is very ill-conditioned (see the numerical section for an illustration of this phenomenon).

The contribution of this work is to propose a Newton's method tailored for efficiently solving the nonlinear system to solve the nonlinear system \cref{eq:obj}.  Specifically, we demonstrate that

\begin{enumerate}

\item Starting from a problem-independent initial guess proposed in Ref.~\cite{DongMengWhaleyEtAl2021}, Newton's method can converge rapidly in all parameter regimes, using standard double precision arithmetic operations. 

\item The computation of the Jacobian matrix can leverage the matrix product states structure of QSP. Notably, the computational cost associated with computing the Jacobian matrix is comparable to that of a \textit{single function evaluation}.

\item The prefactor of the numerical method can be further enhanced by reformulating the symmetric QSP using real-number arithmetics.

\end{enumerate}

We have conducted extensive numerical tests, which have consistently demonstrated the efficiency and robustness of the method. Thus far we have not encountered any instances where the method fails. We have implemented the method in the QSPPACK software package~\footnote{The examples are available on the website \url{https://qsppack.gitbook.io/qsppack/} and the codes are open-sourced in \url{https://github.com/qsppack/QSPPACK}.}.

\vspace{1em}
\noindent\textbf{Related works:}

The phase-factor evaluation was originally conceived to be a challenging task~\cite{LowChuang2017,ChildsMaslovNamEtAl2018}. In the past few years, significant progress has been made to develop efficient algorithms to find phase factors.
These algorithms fall into two categories: factorization methods~\cite{GilyenSuLowEtAl2019,Haah2019,ChaoDingGilyenEtAl2020,Ying2022}, and iterative methods~\cite{DongMengWhaleyEtAl2021,DongLinNiEtAl2022}.

For a given \textit{real} polynomial $f(x)$, factorization methods construct phase factors from the roots of $1-f^2(x)$ in the complex plane, and the roots must be obtained at high precision. As a result, as the polynomial $d$ increases, direct implementation of factorization-based methods is not numerically stable and requires $\Or(d\log(d/\epsilon))$ bits of precision~\cite{GilyenSuLowEtAl2019,Haah2019} ($\epsilon$ is the target accuracy). There have been two recent improvements of factorization-based methods: the capitalization method~\cite{ChaoDingGilyenEtAl2020}, and the Prony method~\cite{Ying2022}. Empirical results indicate that both methods are numerically stable and are applicable to large degree polynomials. Furthermore, the performance of factorization-based methods does not deteriorate near the fully-coherent regime.

Compared to the elaborate construction of factorization-based methods, iterative methods are intuitive, numerically stable, and easy to implement. The idea is to directly tackle the nonlinear system \eqref{eq:obj}, or the equivalent optimization formulation
\begin{equation}\label{eqn:optimization}
    \Psi^* = \argmin_{\Psi} \norm{F(\Psi) - c}_2^2.
\end{equation}

However, due to the complex energy landscape~\cite{WangDongLin2022}, direct optimization from random initial guesses can easily get stuck at local minima and can only be used for low degree polynomials. Ref.~\cite{DongMengWhaleyEtAl2021,WangDongLin2022} propose and study the \emph{symmetric} QSP where the set of phase factors are subjected to a symmetry condition that reduces the degrees of freedom. Ref.~\cite{DongMengWhaleyEtAl2021} further observes that starting from a carefully chosen but problem-independent initial guess, standard optimization methods such as the LBFGS method~\cite{NocedalWright1999} can be robust and stable and can be applied to very high degree polynomials.  Recently, we propose the fixed point iteration (FPI) algorithm that directly tackles the nonlinear system~\cref{eq:obj} and show that the symmetric phase factors have a well-defined limit as the polynomial degree increases towards infinity when the polynomial approaches a smooth (non-polynomial) function. However, it is important to note that in many examples near the fully-coherent regime, the assumptions of those theoretical results are violated. Consequently, gradient-based optimization methods and the FPI method may exhibit slow convergence or fail to converge altogether. 
We would like to remark that the symmetric condition is important for extending QSP to quantum eigenvalue transformation of unitaries (QETU)~\cite{DongLinTong2022}. Additionally, the existing factorization methods are not compatible with the symmetry condition of the phase factors.

\vspace{1em}
\noindent\textbf{Organization:}

The paper is organized as follows. The preliminaries are given in \cref{sec:problem_and_main_alg}. In \cref{sec:problem}, we review relevant concepts in QSP with symmetric phase factors. Then, in \cref{sec:bottleneck}, we discuss the bottleneck towards the fully-coherent regime and also review the iterative methods for finding phase factors in the literature. The matrix product state and its relevance in the structure of QSP are presented in \cref{sec:TT}. Our main algorithm and its implementation are given in \cref{sec:main_algorithm}, where we also discuss the acceleration of the algorithm by leveraging the structure of the problem and a real-number arithmetic representation of QSP. Finally, in \cref{sec:numerics}, we demonstrate our algorithm by presenting the results of numerical experiments.

\section{Preliminaries}\label{sec:problem_and_main_alg}
\subsection{Quantum signal processing with symmetric phase factors}\label{sec:problem}
Quantum signal processing (QSP) represents a class of polynomials in terms of $\mathrm{SU}(2)$ matrices, which is parameterized by phase factors~\cite[Theorem 4]{GilyenSuLowEtAl2019}. The phase factors $\Psi = (\psi_0, \psi_1, \cdots, \psi_d) \in \RR^{d + 1}$ are symmetric if they satisfy the constraint $\psi_i = \psi_{d - i}$ for any $i = 0, \cdots, d$. 
Ref.~\cite{WangDongLin2022} proposes a variant of QSP representation focusing on symmetric phase factors:

\begin{theorem}[\textbf{Quantum signal processing with symmetric phase factors} {\cite[Theorem 1]{WangDongLin2022}}]\label{thm:sym_qsp}
Consider any $P\in \mathbb{C}[x]$ and $Q\in \mathbb{R}[x]$ satisfying the following conditions:
\begin{enumerate}
    \item $\deg(P)= d$ and $\deg(Q)= d-1$,
    \item $P$ has parity $(d \bmod 2)$ and $Q$ has parity $(d-1 \bmod 2)$,
    \item (Normalization condition) $\forall x\in[-1,1]: |P(x)|^2+(1-x^2)|Q(x)|^2=1$,
    \item \label{itm:4} If $d$ is odd, then the leading coefficient of $Q$ is positive.
\end{enumerate}
There exists a unique set of symmetric phase factors $\Psi:=(\psi_0,\psi_1,\cdots,\psi_d)\in D_d$ such that 
\begin{equation}\label{eq:UPQ}
U(x,\Psi)= e^{\I \psi_0 Z} \prod_{j=1}^{d} \left[ W(x) e^{\I \psi_j Z} \right] =\begin{pmatrix}
P(x) & \I Q(x)\sqrt{1-x^2}\\
\I Q(x) \sqrt{1-x^2} & P^* (x)
\end{pmatrix},
\end{equation}
where
\begin{equation}
    D_d=\begin{cases}
    [-\frac{\pi}{2},\frac{\pi}{2})^{\frac{d}{2}} \times [-\pi,\pi) \times [-\frac{\pi}{2},\frac{\pi}{2})^{\frac{d}{2}}, & d \mbox{ is even,}\\
    [-\frac{\pi}{2},\frac{\pi}{2})^{d+1}, &d \mbox{ is odd.}\\
    \end{cases}
\end{equation}
\end{theorem}

In the above equations, $X$ and $Z$ denote Pauli matrices, and $P^*(x)$ represents the complex conjugate of the complex polynomial $P(x)$ obtained by conjugating all its coefficients.

In most applications, only either the real or the imaginary part of the complex polynomial $P(x)=\langle 0|U(x,\Psi)|0\rangle$ is relevant. It can be shown that these two parts can be exchanged by conjugating the unitary matrix product with $\pi/4$-$Z$ rotation, that is,
\begin{equation}\label{eq:re_im_equiv}
    \Re[\langle0|U(x,\Psi)|0\rangle]=\Im\left[\langle 0|e^{\I \frac{\pi}{4}Z}U(x,\Psi)e^{\I \frac{\pi}{4}Z}|0\rangle\right].
\end{equation}
This is equivalent to shifting the edge phase factors $\psi_0, \psi_d$ by $\pi/4$. For the simplicity of presentation, this paper assumes that the imaginary part is relevant. Furthermore, we denote it as $g(x,\Psi)$,
\begin{equation}\label{eqn:qsp-polynomial}
    g(x,\Psi):=\Im[\langle 0|U(x,\Psi)|0\rangle].
\end{equation}
Given any symmetric phase factors $\Psi:=(\psi_0,\psi_1,\ldots,\psi_d)$ of length $d+1$, we define the \emph{reduced phase factors} $\Phi$ as
\begin{equation}
    \Phi=(\phi_0,\phi_1,\ldots,\phi_{\wt{d}-1}):=\begin{cases}
    (\frac{1}{2}\psi_{\wt{d}-1}, \psi_{\wt{d}},\cdots,\psi_{d}), & d \text{ is even},\\
    (\psi_{\wt{d}}, \psi_{\wt{d}-1},\cdots,\psi_{d}), & d \text{ is odd},\\
    \end{cases}
\end{equation}
where $\wt{d}:=\lceil \frac{d+1}{2} \rceil$. For the sake of simplicity, we do not explicitly distinguish the full set of phase factors and the reduced phase factors when they are used as the argument of some function. For example, $U(x,\Phi)$ and $g(x,\Phi)$ are assumed to represent evaluations with respect to the full set of phase factors, $U(x,\Psi)$ and $g(x,\Psi)$. To be embedded in a $\mathrm{SU}(2)$ matrix, it is naturally required that $g(x, \Psi) \le 1$ for any $x \in [-1, 1]$. Hence, the target function $f:\RR\to \RR$ is also normalized so that its norm is bounded:
\begin{equation}
\norm{f}_{\infty}=\max_{x\in[-1,1]} \abs{f(x)}\le 1.
\end{equation}

\cref{thm:sym_qsp} implies that if the target polynomial $f$ of definite parity can be represented as symmetric QSP, it admits a Chebyshev polynomial expansion:
\begin{equation}
f(x)=\begin{cases}
\sum_{j=0}^{\wt{d}-1} c_j T_{2j}(x),& f \mbox{ is even},\\
\sum_{j=0}^{\wt{d}-1} c_j T_{2j+1}(x), & f \mbox{ is odd}.
\end{cases}
\label{eqn:f_expand}
\end{equation}
Let $\mc{F}$ denote the linear mapping from a target polynomial to its Chebyshev-coefficient vector
\begin{equation}
    c:=(c_0, c_1,\cdots,c_{\wt{d}-1})\in\RR^{\wt{d}}.
\end{equation}
This induces the mapping from the set of reduced phase factors to the Chebyshev-coefficient vector of $g(x,\Phi)$,
\begin{equation}\label{eq:F_dfn}
F:\RR^{\wt{d}}\to \RR^{\wt{d}}, \quad F(\Phi):= \mc{F}(g(x,\Phi)).
\end{equation}

\subsection{Iterative methods for finding phase factors and numerical difficulties near the fully-coherent regime}\label{sec:bottleneck}

The QSP problem can also be solved using numerical optimization  
\begin{equation}\label{eqn:optimization_detail}
    \Phi^* = \argmin_{\Phi} \norm{F(\Phi) - c}_2^2 = \argmin_{\Phi} \sum_{j = 1}^{\wt{d}} \abs{g(x_j, \Phi) - f(x_j)}^2.
\end{equation}
Here, $x_j = \cos\left(\frac{(2j - 1)\pi}{4 \wt{d} + 1}\right)$ is the $j$-th node of the Chebyshev polynomial $T_{2 \wt{d}}(x)$. The equality in the optimization problem follows the discrete orthogonality on Chebyshev nodes. In Ref.~\cite{DongMengWhaleyEtAl2021}, the optimization problem is first solved using the LBFGS method and the running complexity is numerically studied. The authors also propose the use of an initial guess $\Phi^0 = (0, 0, \cdots, 0)$, from which the convergence of the LBFGS method is numerically observed to be fast and stable. We remark that the initial guess in the original paper is not identical to this form, due to the difference in the definition. The original paper considers the real part of $\langle 0|U(x,\Psi)|0\rangle$ to encode the polynomial of interest, whereas this paper considers the imaginary part, with equivalence established through \cref{eq:re_im_equiv}. The choice of the initial guess is justified in the theoretical analysis in Ref.~\cite{WangDongLin2022}, which is credited to a class of optima called the maximal solution. In Ref.~\cite{WangDongLin2022}, the authors analyze the energy landscape of the optimization problem and conclude that when the target function is scaled as $\norm{f}_\infty = \Or(1/d)$, the optimization-based algorithm converges locally at $\Or(d^2 \log(1/\epsilon))$ computational cost.

For the target function near the fully-coherent regime, it is hard to guarantee the convergence of optimization-based methods. The ill-conditioned Hessian matrix around the fully-coherent regime poses a challenge for the optimizer to be convergent. The numerical study in Ref.~\cite{DongMengWhaleyEtAl2021} shows that the condition number of the Hessian matrix at the optimum grows rapidly as the target function gets closer to the fully-coherent regime. Furthermore, the theoretical analysis of the optimization landscape also suggests that the region of convergence shrinks as the target function approaches the fully-coherent regime. Hence, the convergence guarantee of optimization-based methods is compromised near the fully-coherent regime.

Using the mapping $F$ defined in \cref{eq:F_dfn}, the problem of finding phase factors can be formulated as solving a nonlinear equation given by \cref{eq:obj}. In Ref.~\cite{DongLinNiEtAl2022}, a fixed-point iteration method (FPI) is proposed for solving \cref{eq:obj}:
\begin{equation}\label{eq:FPI_update}
\Phi^{0} = \mathbf{0}\in\RR^{\wt{d}},\quad\Phi^{t+1}=\Phi^{t}-\frac{1}{2}\left( F\left(\Phi^{t}\right)-c\right).
\end{equation}
Notably, the initial guess of the FPI method coincides with that used in the LBFGS method. The analysis in Ref.~\cite{DongLinNiEtAl2022} demonstrates that the FPI method exhibits linear convergence to the exact solution when $\norm{c}_1\le 0.861$. This result is based on the observation that the update rule acts as a contraction mapping in a neighborhood of the initial guess $\Phi^{0} =\mathbf{0}$. However, this property does not hold universally across the entire domain. The analysis in Ref.~\cite{DongLinNiEtAl2022} reveals that the contraction property is valid when the Chebyshev coefficient vector of the target function lies within an $\ell^1$ ball centered at the origin. In cases where the target function contains significant ``high-frequency'' components, the increasingly large Chebyshev coefficient vector may hinder the contraction of the update rule in \cref{eq:FPI_update}. This situation commonly occurs in various applications; for instance, problematic convergence issues can arise when dealing with functions $\sin(\tau x)$ and $\cos(\tau x)$, where $\tau \gg 1$ is large.

\begin{figure}[htbp]
        \centering
        \includegraphics[width =0.5\textwidth]{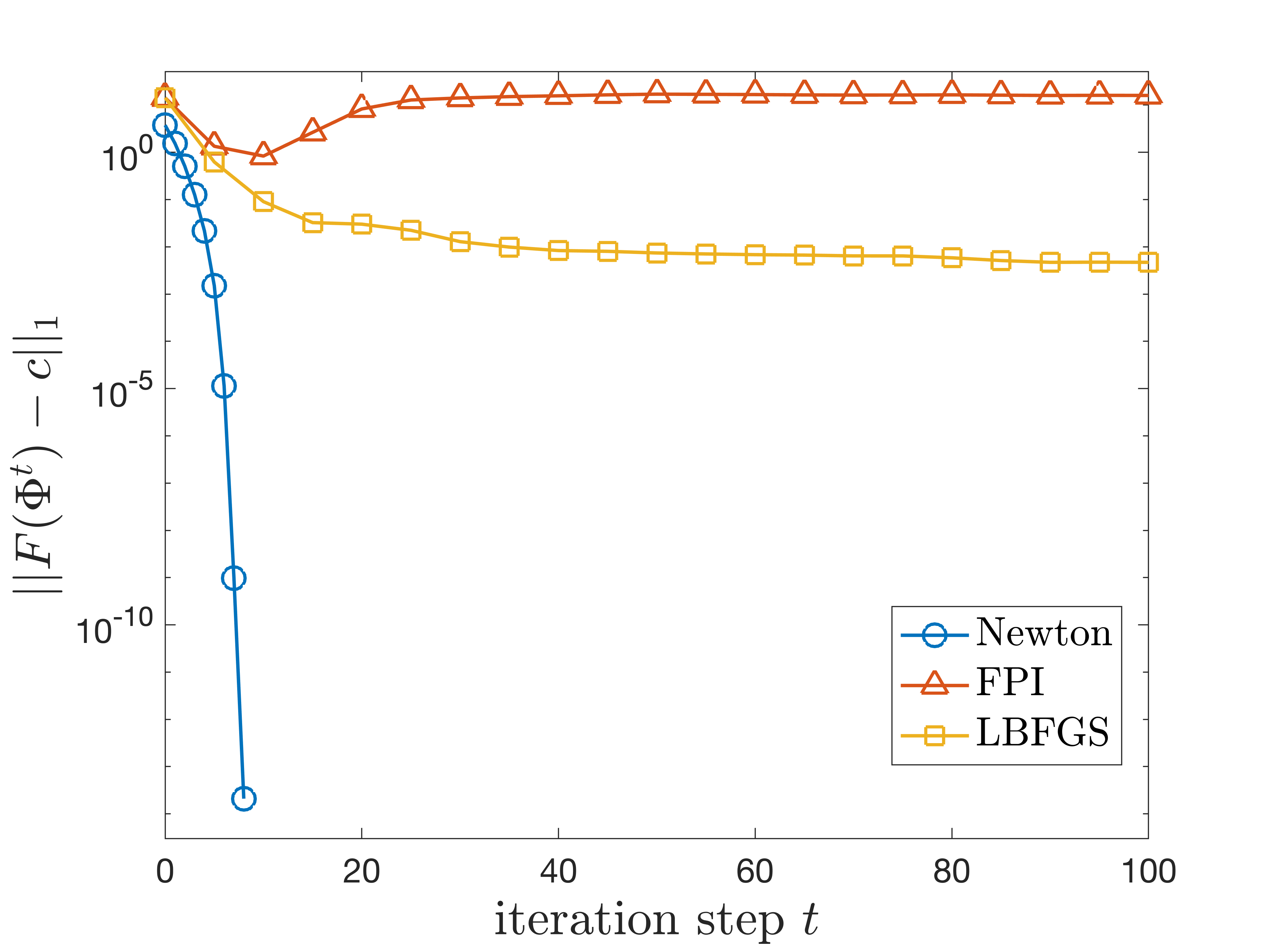}
        \caption{The residual error after each iteration using three different methods to determine phase factors for the target function $f(x) = 0.999\cos(500 x)$ (up to the first $100$ iterations). The stopping criterion is that the residual error reaches below $10^{-13}$. Newton's method converges after 9 steps. The LBFGS method can eventually converge but takes over 1000 steps. The FPI method fails to converge. }\label{fig:fail}
\end{figure}

To conclude the discussion on the challenges faced by iterative methods in the fully-coherent regime, we present a numerical result that substantiates these difficulties. We consider the target function to be a degree-$733$ polynomial approximating $f(x) = 0.999\cos(500 x)$ obtained by truncating the Chebyshev expansion. This function violates the convergence analysis of iterative methods, as discussed earlier. In \cref{fig:fail}, we plot the residual error at each iteration step. The FPI method does not converge at all, while the LBFGS method eventually reaches the optimum. However, the optimizer becomes trapped after the 100th step and requires over 1000 iterations to converge. In contrast, Newton's method, proposed in this paper, exhibits fast and stable convergence in the numerical results. The residual error decreases super-exponentially, consistent with the expected quadratic convergence described in standard textbooks.

 \subsection{Matrix product state structure of quantum signal processing}\label{sec:TT}
The QSP problem, being a well-structured product of $\mathrm{SU}(2)$ matrices, possesses inherent properties that allow for a special tensor structure known as a matrix product state (MPS) or tensor train (TT). These properties can be effectively leveraged to accelerate our numerical algorithm. By exploiting this tensor structure, we achieve a significant reduction in computation complexity, enabling the scalability of the solver for larger-scale applications.

In this subsection, we present a concise overview of the theory and construction of MPS/TT. Additionally, we establish its connection with our problem. To be specific, QSP admits an MPS/TT structure with bond dimension 2 due to its $\mathrm{SU}(2)$-product defining equation \cref{eq:UPQ}.

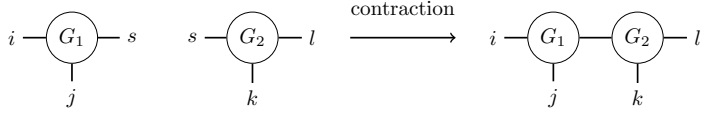
\begin{figure}
    \centering
    \scalebox{0.8}{\begin{tikzpicture}
    \node(v0) at (0,0) {$i$};
    \node[draw, shape=circle] (v1) at (1,0) {$G_1$};
    \node(v2) at (2,0) {$s$};
    \node(v3) at (3,0) {$s$};
    \node[draw, shape=circle] (v4) at (4,0) {$G_2$};
    \node(v5) at (5,0) {$l$};
    \node(v6) at (5.5,0) {};
    \node(v7) at (7.5,0) {};
    \node(v8) at (8,0) {$i$};
    \node[draw, shape=circle] (v9) at (9,0) {$G_1$};
    \node[draw, shape=circle] (v10) at (10.4,0) {$G_2$};
    \node(v11) at (11.4,0) {$l$};
    \node (u1) at (1,-1) {$j$};
    \node (u4) at (4,-1) {$k$};
    \node (u6) at (6.5,0.5) {contraction};
    \node (u9) at (9,-1) {$j$};
    \node (u10) at (10.4,-1) {$k$};
    \draw [thick, ->] (v6) -- (v7);
    \draw [thick] (v0) -- (v1) -- (v2)  (v3) -- (v4) -- (v5) (v8) -- (v9) -- (v10) -- (v11)
    (v1) -- (u1)
    (v4) -- (u4)
    (v9) -- (u9)
    (v10) -- (u10);
    \end{tikzpicture}}
    \caption{A graphical visualization of two order-3 tensors and their contraction.}
    \label{fig:visual3tensor}
\end{figure}

 Given a field $\FF = \RR \text{ or } \CC$, an \emph{order-$r$ tensor} is referred to as a multidimensional array $G \in \FF^{n_1 \times \cdots \times n_r}$ where the $r$-tuple $(n_1, \cdots, n_r) \in \NN^r$ specifies the size of the tensor. Each entry of the tensor can be accessed with multi-indices, $G(i_1, \cdots, i_r)$, where $1 \le i_j \le n_j, \forall j = 1, \cdots, r$. The contraction of two tensors yields a new tensor by summing over the specified indices. For example, if $G_1 \in \FF^{n_1\times n_2\times n_3}, G_2 \in \FF^{m_1\times m_2\times n_3}$ are two order-$3$ tensors, $G_3(i,j,k,l) := \sum_{s=1}^{n_3} G_1(i,j,s) G_2(k,l,s)$ element-wisely defines an order-$4$ tensor $G_3 \in \FF^{n_1\times n_2\times m_1\times m_2}$ by contracting the index $s$. A graphical illustration is given in \cref{fig:visual3tensor}. Specifically, the contraction of order-$2$ tensors (i.e. matrices) coincides with matrix multiplication.

 A (parametric) tensor $G(\alpha) \in \FF^{n_1 \times \cdots \times n_r}$ is called an MPS or TT if each of its entries can be expressed as a product of matrices~\cite{Oseledets2011}. To be precise, an MPS/TT can be written as
 $$G(i_1, \cdots, i_r; \alpha) = \mc{G}_1(i_1; \alpha) \mc{G}_2(i_2; \alpha) \cdots \mc{G}_r(i_r; \alpha).$$
In the above expression, each $\mc{G}_j(\alpha)$ is an order-$3$ tensor. By fixing an index $i_j$, $\mc{G}_j(i_j; \alpha) = [G_j(k,l,i_j; \alpha)]_{kl} \in \FF^{m_{j-1} \times m_j}$ becomes order-$2$ which is equivalent to a matrix of size $(m_{j-1}, m_j)$. The dangling index $i_j$ is referred to as the mode index (or external index). The indices $k, l$ which are dummy in contraction are referred to as rank core indices. The right-hand side of the defining equation is a shorthand for contracting all non-fixed indices by matrix multiplication. The contracted tensor $G(i_1, \cdots, i_r; \alpha)$ is assumed to be a scalar entry-wisely. Hence, the dangling components $\mc{G}_1$ and $\mc{G}_r$ are set to order-$2$ tensors, namely, $m_0 = m_r = 1$. The bond dimension of an MPS/TT is defined to be the maximal dimension in the contraction $\max_{0 \le j \le r} m_j$.

\begin{figure}[tbhp]
\centering
\begin{tikzpicture}
    \node[draw, shape=circle] (v0) at (0,0) {$\mc{R}_0$};
    \node[draw, shape=circle] (v1) at (1,0) {$\mc{W}$};
    \node[draw, shape=circle] (v2) at (2,0) {$\mc{R}$};
    \node(v3) at (3,0) {$\cdots$};
    \node[draw, shape=circle] (v4) at (4,0) {$\mc{W}$};
    \node[draw, shape=circle] (v5) at (5,0) {$\mc{R}$};
    \node[draw, shape=circle] (v6) at (6,0) {$\mc{W}$};
    \node(v7) at (7,0) {$\cdots$};
    \node[draw, shape=circle] (v8) at (8,0) {$\mc{R}$};
    \node[draw, shape=circle] (v9) at (9,0) {$\mc{W}$};
    \node[draw, shape=circle] (v10) at (10,0) {$\mc{R}_d$};
    \node (u0) at (0,-1) {$\psi_0$};
    \node (u1) at (1,-1) {$x$};
    \node (u2) at (2,-1) {$\psi_1$};
    \node (u4) at (4,-1) {$x$};
    \node (u5) at (5,-1) {$\psi_i$};
    \node (u6) at (6,-1) {$x$};
    \node (u8) at (8,-1) {$\psi_{d-1}$};
    \node (u9) at (9,-1) {$x$};
    \node (u10) at (10,-1) {$\psi_d$};
    \draw [thick] (v0) -- (v1) -- (v2) -- (v3) -- (v4) -- (v5) -- (v6) -- (v7) -- (v8) -- (v9) -- (v10)
    (v0) -- (u0)
    (v1) -- (u1)
    (v2) -- (u2)
    (v4) -- (u4)
    (v5) -- (u5)
    (v6) -- (u6)
    (v8) -- (u8)
    (v9) -- (u9)
    (v10) -- (u10);
\end{tikzpicture}
\caption{A graphical visualization of $\langle 0 | U(x, \Psi) | 0 \rangle$ as a MPS/TT structure (of bond dimension $2$).}
\label{fig:visualfull}
\end{figure}
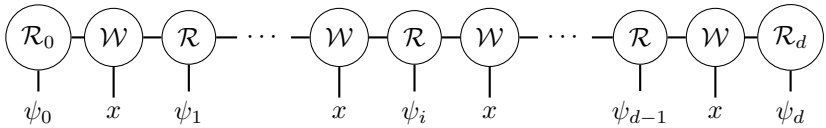

Using the language of tensors, the upper-left entry of the QSP unitary matrix defined in \cref{eq:UPQ} can be interpreted as an MPS/TT of bond dimension $2$. To see this, assuming that $x$ and a full set of phase factors $\Psi := (\psi_0, \cdots, \psi_d)$ are given, the QSP matrix entry of interest is
\begin{equation}\label{eq:MPS}
    \langle 0 | U(x, \Psi) | 0 \rangle = \mc{R}_0(\psi_0) \mc{W}(x) \mc{R}(\psi_1) \mc{W}(x) \cdots \mc{W}(x) \mc{R}(\psi_{d-1}) \mc{W}(x) \mc{R}_d(\psi_d) \in \CC.
\end{equation}
Here, the boundary components $\mc{R}_0(\psi_0) := (e^{\I\psi_0}, 0), \mc{R}_d(\psi_d) := (e^{\I\psi_d},0)^\top$ are two-dimensional complex vectors. Furthermore, $\mc{W}(x) := e^{\I\arccos(x) X}$ and $\mc{R}(\psi_j) := e^{\I\psi_jZ}$ are $2$-by-$2$ complex matrices. By identifying $x,\psi$ as external indices, the graphical visualization of this interpretation is given in \cref{fig:visualfull}.

\section{Newton's method} \label{sec:main_algorithm}
When using iterative algorithms discussed in the previous section to find phase factors, the issue of convergence becomes increasingly significant when the target function is close to the fully-coherent regime. To remedy these difficulties, we propose using Newton's method to solve the nonlinear equation \cref{eq:obj} for phase-factor evaluation. In this section, we introduce this method and discuss its implementation. The core techniques to accelerate the algorithm are fast Jacobian evaluation based on the MPS/TT structure and a real-arithmetic formalism of symmetric QSP.

Newton's method can be viewed as an improvement over the FPI method by taking the local landscape into account. It can be verified that the Jacobian of the nonlinear equation in \cref{eq:obj} at the origin coincides with a doubled identity matrix, that is, $DF(\mathbf{0}) = 2 \mathbf{I}$. Hence, the FPI method is a variant of Newton's method, where the Jacobian is approximated along the iteration by that at the initial point, which is the origin. To be precise, the update rules $\Phi^{t+1} = T(\Phi^t)$ of both methods are written as
\begin{equation}
\begin{split}
    T_\mathrm{Newton}(\Phi) &= \Phi - DF(\Phi)^{-1} (F(\Phi) - c),\\
    \text{and } T_\mathrm{FPI}(\Phi) &= \Phi - DF(\mathbf{0})^{-1} (F(\Phi) - c), \text{ where } DF(\mathbf{0}) = 2 \mathbf{I}.
\end{split}
\end{equation}
The algorithm based on the first update rule is outlined in \cref{alg:newton}. In the remainder of this section, we will discuss an accelerated implementation of this algorithm leveraging the structure of the symmetric QSP problem.
\begin{algorithm}[htbp]
\caption{Newton's method for finding reduced phase factors}
\label{alg:newton}
\begin{algorithmic}
\STATE{\textbf{Input:}  Chebyshev-coefficient vector $c$ of a target polynomial, and stopping criteria.}
\STATE{Initiate $t=0$ and $\Phi^{0} $ to be zero vector $\mathbf{0}$;}
\WHILE{stopping criterion is not satisfied}
        \STATE{Compute $DF(\Phi^{t})$;}
        \STATE{Update $\Phi^{t+1}=\Phi^{t}-DF(\Phi^{t})^{-1}\left( F\left(\Phi^{t}\right)-c\right)$;}
        \STATE{Set $t=t+1$;
                }
\ENDWHILE
\STATE{\textbf{Output:} Reduced phase factors $\Phi^{t}$.}
\end{algorithmic}
\end{algorithm}

\subsection{Jacobian of the problem}
The update rule of Newton's method utilizes the Jacobian matrix of  $F(\Phi)$, which is denoted as $DF(\Phi)$. According to the defining equation \cref{eq:F_dfn} of $F$, the $i$-th column of $DF(\Phi)$ is 
\begin{equation}\label{eqn:DF_column}
    \frac{\partial F(\Phi)}{\partial \phi_i}= \mc{F}\left(\frac{\partial g(x,\Phi)}{\partial \phi_i}\right) \in \mathbb{R}^{\wt{d}},\ i=0,\cdots,\wt{d}-1.
\end{equation}
A straightforward approach for constructing the Jacobian matrix is to compute it column-wise without any optimization. This method involves performing the following procedure independently for each $0 \le i < \wt{d}$: evaluating $\partial g(x_k, \Phi) / \partial \phi_i$ at approximately $\Or(d)$ distinct points and then applying a discrete Fourier transformation. Each evaluation of $\partial g(x_k, \Phi) / \partial \phi_i$ requires $\Or(d)$ multiplications of $\mathrm{SU}(2)$. Consequently, the complexity for computing a column of the Jacobian is $\Or(d^2)$. As a result, the overall complexity of this Jacobian evaluation is $\Or(d^3)$. It is important to note that this approach does not take into account the structural characteristics of the problem, leaving room for potential optimization strategies.

In the subsequent subsection, we will present an accelerated evaluation method that capitalizes on the MPS/TT structure of the problem. This improved approach leads to a notable reduction in the complexity of Jacobian evaluation, from $\Or(d^3)$ to $\Or(d^2 \log(d))$.

In the remainder of this subsection, we delve into the structure of the Jacobian matrix columns. For the sake of simplicity, we assume that $d$ is even. As a reminder, in the even case, the full set of phase factors can be represented as $\Psi = (\phi_{\wt{d}-1}, \cdots,\phi_1, 2\phi_0,\phi_1, \cdots, \phi_{\wt{d}-1})$. This choice does not lose generality, as a similar derivation can be obtained for the odd case.

We first observe that $\frac{\partial g(x,\Phi)}{\partial \phi_i} = \Im\left[ \braket{0| \frac{\partial}{\partial \phi_i} U(x, \Psi) |0} \right]$, and that taking derivative on the unitary matrix $U(x, \Psi)$ is equivalent to the insertion of an additional $\I Z = e^{\I \pi Z / 2}$ in the matrix product. Due to symmetry, the derivative leads to two matrix products with insertion. Specifically, we have: 
\begin{equation}
    \frac{\partial}{\partial \phi_i} U(x, \Psi) = U(x, \Psi + \frac{\pi}{2} e_{\wt{d}-1-i}) +  U(x, \Psi + \frac{\pi}{2} e_{\wt{d}-1+i}),\quad \forall 0\leq i \leq \wt{d}-1.
\end{equation}
We remark that $\Psi + \frac{\pi}{2} e_{\wt{d}-1+i}$ is the reversed ordering of $\Psi +\frac{\pi}{2} e_{\wt{d}-1-i}$, which is helpful for the simplification. Consequently, $U(x, \Psi + \frac{\pi}{2} e_{\wt{d}-1+i})$ is identical to the transpose of $U(x, \Psi + \frac{\pi}{2} e_{\wt{d}-1-i})$ since the transpose reverses the order due to the symmetry of matrix $Z$. Hence
\begin{equation}
    \begin{split}
        \frac{\partial g(x,\Phi)}{\partial \phi_i} &= \Im\left[ \braket{0| \frac{\partial}{\partial \phi_i} U(x, \Phi) |0} \right] \\
    &= \Im[\braket{0|U(x,\Psi + \frac{\pi}{2} e_{\wt{d}-1-i})|0}] + \Im[\braket{0|U (x,\Psi + \frac{\pi}{2} e_{\wt{d}-1+i})|0}]\\
    &= \Im[\braket{0|U(x,\Psi + \frac{\pi}{2} e_{\wt{d}-1-i})|0}] + \Im[\braket{0|U(x,\Psi + \frac{\pi}{2} e_{\wt{d}-1-i})^\top |0}]\\
    &= 2\Im[\braket{0|U(x,\Psi+\frac{\pi}{2} e_{\wt{d}-1-i}))|0}] = 2 g(x,\Psi + \frac{\pi}{2}e_{\wt{d}-1-i}).
    \end{split}
\end{equation}
While $\Psi+\frac{\pi}{2} e_{\wt{d}-1-i}$ is not symmetric, the evaluation of the induced polynomial is still well defined. To extract its Chebyshev coefficients, it suffices to sample this polynomial on the Chebyshev nodes $\{x_k = \cos(2\pi k / (2d + 1)): k = 0, \cdots, d\}$. Subsequently, the Chebyshev coefficients can be extracted from the sample by performing Fast Fourier Transformation (FFT). The detail is given in \cref{alg:F_eval}.
 
\begin{algorithm}[htbp]
\caption{Compute $\mc{F}(g(x,\Psi^\sharp))$.}
\label{alg:F_eval}
\begin{algorithmic}
\STATE{\textbf{Input:} A full set of phase factors $\Psi^\sharp$ of length $d+1$ ($\Psi^\sharp$ is not necessarily symmetric).}
\STATE{Initialize $\vg = (0, 0, \cdots) \in \RR^{2d+1}$.}
\STATE{Evaluate $\vg_j \leftarrow g(x_j, \Psi^\sharp),x_j=\cos\left(\frac{2\pi j}{2d+1}\right),j = 0, \cdots, d$.}
\STATE{Evaluate $\vg_j \leftarrow \vg_{2d+1-j}$, $j=d+1,\cdots, 2d$.}
\STATE{Compute $\vv_l\leftarrow \Re\left(\sum_{j=0}^{2d-1} \vg_j e^{-\I \frac{2\pi}{2d+1}jl}\right),l=0,\ldots,d$ using FFT.}
\IF{$(d \text{ mod } 2) = 0$}
\STATE{$\mc{F}(g(x,\Psi^\sharp))\leftarrow \frac{2}{2d+1}(\frac{\vv_0}{2}, \vv_2, \vv_4,\cdots, \vv_d)$.}
\ELSE
\STATE{$\mc{F}(g(x,\Psi^\sharp))\leftarrow \frac{2}{2d+1}(\vv_1, \vv_3, \vv_5,\cdots, \vv_d)$.}
\ENDIF
\STATE{\textbf{Output:} $\mc{F}(g(x,\Psi^\sharp))$.}
\end{algorithmic}
\end{algorithm}

\subsection{Efficient evaluation of the Jacobian matrix}\label{sec:MPS}

In practice, The evaluation of the Jacobian matrix $DF(\Phi)$ constitutes the most computationally demanding step in Newton's method. In this subsection, we propose an efficient approach to compute the Jacobian matrix, taking advantage of the MPS/TT structure of the problem. By doing so, we can significantly reduce the overall computational complexity. As we will illustrate, the computation of different columns of the Jacobian matrix exhibits substantial overlap. This indicates that we can reuse intermediate computational results and avoid redundancy, leading to increased efficiency.

\begin{figure}[tbhp]
\centering
\scalebox{0.8}{\begin{tikzpicture}[decoration={brace,mirror,amplitude=7}]
    \node[draw, shape=circle] (v0) at (0,0) {$\mc{R}_0$};
    \node[draw, shape=circle] (v1) at (1,0) {$\mc{W}$};
    \node[draw, shape=circle] (v2) at (2,0) {$\mc{R}$};
    \node(v3) at (3,0) {$\cdots$};
    \node[draw, shape=circle] (v4) at (4,0) {$\mc{W}$};
    \node[draw, shape=circle] (v5) at (5,0) {$\mc{R}$};
    \node[draw, shape=circle] (v6) at (6,0) {$\mc{W}$};
    \node(v7) at (7,0) {$\cdots$};
    \node[draw, shape=circle] (v8) at (8,0) {$\mc{W}$};
    \node[draw, shape=circle] (v9) at (9,0) {$\mc{R}$};
    \node[draw, shape=circle] (v10) at (10,0) {$\mc{W}$};
    \node(v11) at (11,0) {$\cdots$};
    \node[draw, shape=circle] (v12) at (12,0) {$\mc{R}$};
    \node[draw, shape=circle] (v13) at (13,0) {$\mc{W}$};
    \node[draw, shape=circle] (v14) at (14,0) {$\mc{R}_d$};
    \node (u0) at (0,-1) {$\phi_{\wt{d}-1}$};
    \node (u1) at (1,-1) {$x$};
    \node (u2) at (2,-1) {$\phi_{\wt{d}-2}$};
    \node (u4) at (4,-1) {$x$};
    \node (u5) at (5,-1) {$\phi_i$};
    \node (u6) at (6,-1) {$x$};
    \node (u8) at (8,-1) {$x$};
    \node (u9) at (9,-1) {$2\phi_{0}$};
    \node (u10) at (10,-1) {$x$};
    \node (u12) at (12,-1) {$\phi_{\wt{d}-2}$};
    \node (u13) at (13,-1) {$x$};
    \node (u14) at (14,-1) {$\phi_{\wt{d}-1}$};
    \node (e1) at (4.5,0.6) {};
    \node (e2) at (4.5,-0.6) {};
    \node (e3) at (5.5,0.6) {};
    \node (e4) at (5.5,-0.6) {};
    \node (e5) at (8.5,0.6) {};
    \node (e6) at (8.5,-0.6) {};
    \node (e7) at (9.5,0.6) {};
    \node (e8) at (9.5,-0.6) {};
    \draw [thick] (v0) -- (v1) -- (v2) -- (v3) -- (v4) -- (v5) -- (v6) -- (v7) -- (v8) -- (v9) -- (v10) -- (v11) -- (v12) -- (v13) -- (v14)
    (v0) -- (u0)
    (v1) -- (u1)
    (v2) -- (u2)
    (v4) -- (u4)
    (v5) -- (u5)
    (v6) -- (u6)
    (v8) -- (u8)
    (v9) -- (u9)
    (v10) -- (u10)
    (v12) -- (u12)
    (v13) -- (u13)
    (v14) -- (u14);
    \draw[dashed] (e1) -- (e2) (e3)--(e4) ;
    \draw [decorate] ([yshift=-3mm]u0.west) --node[below=3mm]{$\mc{M}_\text{left}^{(i)}$} ([yshift=-3mm]u4.east);
    \draw [decorate] ([yshift=-3mm]u6.west) --node[below=3mm]{$\mc{M}_\text{right}^{(i)}$} ([yshift=-3mm]u14.east);
\end{tikzpicture}}
\caption{A graphical visualization of the isolation and grouping when evaluating the derivative $\langle 0 | \partial_{\phi_i} U(x, \Phi) | 0 \rangle$.}
\label{fig:visualsym}
\end{figure}
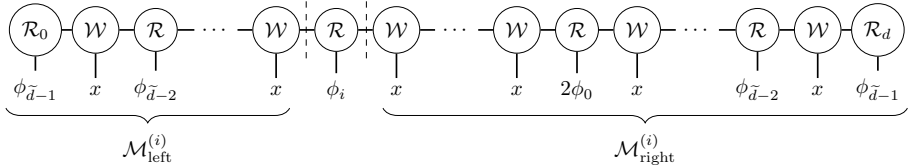

Without loss of generality, we consider the case where $d$ is even in the derivation. The odd case can be analyzed similarly. Recall that the full set of phase factors is defined as 
\begin{equation}
    \Psi := (\psi_0,\psi_1,\cdots, \psi_d)= (\phi_{\wt{d}-1}, \cdots, \phi_1,2\phi_0, \phi_1, \cdots, \phi_{\wt{d}-1}),
\end{equation}
where $\Phi = (\phi_0, \cdots,\phi_{\wt{d}-1})$ are the reduced phases factors.
We observe that each column of the Jacobian matrix is directly associated with taking the derivative of a QSP without symmetry, which arises from the insertion of the $\I Z$ matrix. In the absence of symmetry constraint of phase factors, each phase factor $\phi_i$ is independent. When calculating the derivative with respect to $\phi_i$, we can separate the matrix multiplication into three parts
\begin{equation}
    \langle 0 | U(x, \Psi) | 0 \rangle = \mc{M}_\text{left}^{(i)} e^{\I \phi_i Z} \mc{M}_\text{right}^{(i)},
\end{equation}
where
\begin{equation}
    \begin{split}
        \mc{M}_\text{left}^{(i)} &:= \mc{R}_0(\phi_{\wt{d}-1}) \prod_{j=\wt{d}-2}^{i+1} \left[\mc{W}(x)\mc{R}(\phi_j)\right] \mc{W}(x),\\
        \mc{M}_\text{right}^{(i)} &:= \prod_{j=i-1}^{0} \left[\mc{W}(x)\mc{R}(\phi_j)\right] \prod_{j=0}^{\wt{d}-2} \left[\mc{R}(\phi_j)\mc{W}(x)\right]\mc{R}_d(\phi_{\wt{d}-1}).
    \end{split}
\end{equation}
The left and right components are irrelevant to taking the derivative with respect to $\phi_i$ because
\begin{equation}
     \langle 0 |  \partial_{\phi_i} U(x, \Phi) | 0 \rangle =2 \langle 0 |  U(x, \Psi + \frac{\pi}{2} e_{k-i}) | 0 \rangle = 2\I \mc{M}_\text{left}^{(i)} Z e^{\I \phi_i Z} \mc{M}_\text{right}^{(i)}.
\end{equation}
Consequently, the intermediate quantities $\mc{M}_\text{left}^{(i)}$ and $\mc{M}_\text{right}^{(i)}$ can be stored and maintained in the computation process. \Cref{fig:visualsym} visually illustrates this idea.

Transiting to the next step, the intermediate quantities are updated through matrix multiplications
\begin{equation*}
   \mc{M}_\text{left}^{(i+1)} \leftarrow \mc{M}_\text{left}^{(i)}\mc{W}^{-1}(x) e^{-\I \phi_{i+1} Z},\quad\text{and} \quad \mc{M}_\text{right}^{(i+1)} \leftarrow \mc{W}(x) e^{\I \phi_{i} Z} \mc{M}_\text{right}^{(i)} .
\end{equation*}

By utilizing intermediate quantities, the computation of the derivatives, which are the columns of the Jacobian matrix before FFT, can be performed simultaneously, resulting in a computational cost of $\Or(d^2)$ rather than $\Or(d^3)$ in the straightforward method. The overall complexity of computing the Jacobian matrix is $\mc{O}(d^2\log d)$ due to the use of FFT. The detailed procedure is summarized in Algorithm \ref{alg:DF}.

\begin{algorithm}[htbp]
\caption{Compute Jacobian matrix $DF(\Phi)$ using the MPS structure.}
\label{alg:DF}
\begin{algorithmic}
\STATE{\textbf{Input:} Reduced phase factors $\Phi$ of length $\wt{d}$ and parity.}
\STATE{Set $ d= 2\wt{d}-2 $ + parity and initialize $\vg$ as a zero matrix of size $\wt{d}\times(2d+1)$.}
\FOR{$j = 0,\cdots, d$}
\STATE{Set $x_j=\cos\left(\frac{2\pi j}{2d+1}\right)$.}
\STATE{$\mc{M}_\mathrm{left}(x_j) = (1,0)\prod_{i=\wt{d}-1}^1 \left(e^{\I \phi_i Z} W(x_j)\right)$.}
\STATE{$ \mc{M}_\mathrm{right}(x_j) = e^{\I \phi_0 Z} \mc{M}_\mathrm{left}(x_j)^\top$.}
\IF{parity is odd}
\STATE{$\mc{M}_\mathrm{right}(x_j) = W(x_j)\mc{M}_\mathrm{right}(x_j)$.}
\ENDIF
\STATE{$g_{0,j}\leftarrow 2 \Im[\mc{M}_\mathrm{left} \I Z \mc{M}_\mathrm{right} ].$}
\FOR{$ i= 1, \cdots, \wt{d}-1 $}
\STATE{$\mc{M}_\mathrm{left}(x_j)\leftarrow \mc{M}_\mathrm{left}(x_j)W^{-1}(x_j) e^{-\I \phi_i Z}$.}
\STATE{$\mc{M}_\mathrm{right}(x_j)\leftarrow W(x_j) e^{\I \phi_i Z}\mc{M}_\mathrm{right}(x_j)$.}
\STATE{$g_{i,j}\leftarrow 2 \Im[\mc{M}_\mathrm{left} \I Z \mc{M}_\mathrm{right} ]$.}
\ENDFOR
\ENDFOR
\STATE{Evaluate $\vg_{i,j} \leftarrow \vg_{i ,2d+1-j}$, $j=d+1,\cdots, 2d$.}
\STATE{Compute $\vv_{il}\leftarrow \Re\left(\sum_{j=0}^{2d-1} \vg_{i,j} e^{-\I \frac{2\pi}{2d+1}l j}\right),l=0,\ldots,d$ using FFT.}
\IF{parity = 0}
\STATE{$\frac{\partial F(\Phi)}{\partial \phi_i} \leftarrow \frac{2}{2d+1}(\frac{\vv_{i0}}{2}, \vv_{i2}, \vv_{i4},\cdots, \vv_{id})$.}
\ELSE
\STATE{$\frac{\partial F(\Phi)}{\partial \phi_i} \leftarrow \frac{2}{2d+1}(\vv_{i1}, \vv_{i3}, \vv_{i5},\cdots, \vv_{id})$.}
\ENDIF
\STATE{\textbf{Output:} $D F(\Phi)$.}
\end{algorithmic}
\end{algorithm}

\subsection{Formalism of symmetric QSP in real arithmetic operations}\label{sec:Real}

In the existing literature, the conventional formalism of QSP is typically presented in terms of the product of $\mathrm{SU}(2)$ matrices, which involves complex arithmetic operations. This complex arithmetic formalism is both necessary and sufficient for general QSP, as the induced polynomials $P$ and $Q$ are complex without any additional symmetry constraints. However, in the case of symmetric QSP, according to \cref{thm:sym_qsp}, the polynomial $Q$ is a real polynomial. This observation raises the question of whether the formalism of QSP can be simplified to accommodate this symmetry.

In this subsection, we will introduce a formalism for symmetric QSP that utilizes real arithmetic operations. This alternative formalism not only proves to be beneficial for Newton's method proposed in this paper but also enhances the implementation of other algorithms designed to solve symmetric QSP, resulting in a constant improvement in the prefactor of the overall computational complexity.

The core idea is that $\mathrm{SU}(2)$ is homeomorphic to $\mathbb{S}^3 \subset \mathbb{R}^4$, which arises from the parametric form of general $\mathrm{SU}(2)$ matrices. By imposing the symmetric constraint, the upper-right entry of the consequent $\mathrm{SU}(2)$ matrix is purely imaginary. Consequently, we can associate any symmetric QSP matrix with a vector in $\mathbb{S}^2 \subset \mathbb{R}^3$. The identification is
\begin{equation}
\begin{split}
    & U(x, \Phi) = \begin{pmatrix}
p(x, \Phi) + \I g(x, \Phi) & \I \sqrt{1 - x^2} q(x, \Phi)\\
\I \sqrt{1 - x^2} q(x, \Phi) & p(x, \Phi) - \I g(x, \Phi)
\end{pmatrix} \in \mathrm{SU}(2)\\
\leftrightarrow& V(x, \Phi) := \begin{pmatrix}
    p(x, \Phi), \ g(x, \Phi), \ \sqrt{1 - x^2} q(x, \Phi)
\end{pmatrix}^\top \in \mathbb{S}^2.
\end{split}
\end{equation}

Under the identification we introduced, the matrix multiplication in symmetric QSP is equivalent to interleaved rotations in $\mathrm{SO}(3)$. This relation is quantified by the following recurrence equation:
\begin{equation}
    V(x, (\phi_k, \phi_{k - 1}, \cdots )) = R_z(2\phi_k) R_x(2 \arccos(x)) V(x, (\phi_{k - 1}, \cdots )),
\end{equation}
where the $\mathrm{SO}(3)$ rotations are
\begin{equation}
    R_z(2 \phi) = \begin{pmatrix}
    \cos{2\phi} & -\sin{2\phi} & \\
    \sin{2\phi} & \cos{2\phi} & \\
    & & 1
    \end{pmatrix} \text{ and } R_x(2 \theta) =  \begin{pmatrix}
    \cos(2\theta) & & -\sin(2\theta)\\
     & 1 & \\
     \sin(2\theta) & & \cos(2\theta)
    \end{pmatrix}.
\end{equation}

For further details on this identification, we refer readers to \cref{sec:appendix_QSP_real_arithmetic}. By leveraging this identification, the QSP polynomials can be derived from the product of real matrices, leading to a faster computation with a constant improvement in the prefactor, compared to evaluating them using the product of complex matrices.

\section{Experiments}\label{sec:numerics}
In this section, we demonstrate the performance of Newton's method in solving phase factors through various numerical examples. These examples are essential for solving scientific computing problems using quantum algorithms. We begin by introducing the numerical examples, followed by the presentation and discussion of the numerical results in the rest of this section.

\subsection{Setup of numerical examples}
The core of quantum algorithm design based on QSP lies in the abstraction of the original problem as a matrix function transformation. This transformation allows us to encode the desired function or its polynomial approximation by finding the appropriate phase factors. In order to illustrate this procedure, we present the following examples.

\emph{Quantum Hamiltonian simulation.} The problem of quantum Hamiltonian simulation involves finding an efficient method for implementing the time evolution matrix of a Hamiltonian matrix, denoted as $H \mapsto \exp(- \I \tau H)$, for a given evolution time $\tau$. In Ref.~\cite{LowChuang2017}, a near-optimal quantum Hamiltonian simulation algorithm based on QSP is proposed. This algorithm abstracts the problem into a function approximation task, where the target function $f(x) = e^{- \I \tau x}$ is parametrized using QSP. The Chebyshev series expansion, known as the Jacobi-Anger expansion~\cite{LowChuang2017}, is commonly employed to approximate this target function:
\begin{equation}
        \label{eq;Jacobi-Anger}
        e^{-\I\tau x}=J_0(x)+2\sum_{k\text{ even}}(-1)^{k/2} J_k(\tau) T_k(x) + 2\I \sum_{k \text{ odd}} (-1)^{(k-1)/2} J_k(\tau) T_k(x),
\end{equation}
where $J_k$'s are the Bessel functions of the first kind. As a result, by truncating the Jacobi-Anger series, a polynomial approximation of the target function can be obtained. The real and imaginary parts of the truncated series, which approximate $\cos(\tau x)$ and $\sin(\tau x)$ respectively, serve as the target polynomials for two separate QSP phase-evaluation problems. To ensure that the truncation error is upper-bounded by $\epsilon_0$, it is sufficient to choose the degree of truncation as $d = e|\tau|/2 + \log(1/\epsilon_0)$.

\emph{Quantum Gaussian filter.} The quantum Gaussian filter is a matrix function parameterized by $\mu$ and $\sigma$. It is proportional to $\exp(- (H - \mu I)^2 / \sigma^2)$, where $H$ is the Hamiltonian matrix. This matrix function is designed to localize around the given ``energy level'' $\mu$, with the degree of localization controlled by the bandwidth parameter $\sigma$. The function suppresses eigenvalues of $H$ that are far from $\mu$. Ideally, one would choose $\mu$ to be close to an eigenvalue of $H$, allowing the matrix function to approximate the projection onto the corresponding eigenspace.

The quantum Gaussian filter serves as an intermediate subroutine for near-optimal quantum linear system solvers~\cite{LinTong2020}. However, directly decomposing the defining function of the quantum Gaussian filter may result in exponentially large scaling factors due to hyperbolic functions. To address this issue and improve numerical stability, one can shift and rescale the Hamiltonian so that its eigenvalues lie in a smaller subinterval $D_\kappa = [1/\kappa, 1]$ within the positive half-axis. By employing this eigenvalue shifting technique, it is sufficient to approximate the Gaussian density function in the positive half-axis. Thus, the target function is set to $f(x) = e^{- (|x| - \mu)^2 / \sigma^2}$ as an even extension.

\emph{Heaviside energy filter.} Heaviside function is widely used in classical applications such as signal processing and filter design. It also plays a crucial role as a subroutine in quantum algorithms for ground-state energy estimation and ground state preparation~\cite{DongLinTong2022}.

Consider a Hamiltonian matrix that has been shifted and scaled so that its eigenvalues lie in the interval $[0, 1]$. The Heaviside energy filter $f(H)$ attenuates the high-energy components of the Hamiltonian. The function $f(x)$ is defined as follows:
\begin{equation}
    f(x) = \begin{cases}
        1 &  \abs{x}<0.5\\
        \frac{1}{2} & \abs{x} = 0.5\\
        0 & \abs{x}>0.5
    \end{cases}.
\end{equation}
To address the singularity at $0.5$, we assume that the target function only needs to be approximated within the interval $D_\delta = [0, (1 - \delta)/2] \cup [(1 + \delta) / 2, 1]$. This allows us to focus on the desired energy range and mitigate the effects of the singularity.

\emph{Matrix inversion.} Matrix inversion is a fundamental topic in numerical linear algebra with wide-ranging applications, including numerical optimization and least squares problems. In the context of function transformation, the equivalent problem is to implement the transformation $H \mapsto f(H) = H^{-1}$. If the matrix has a condition number of $\kappa = \mathrm{cond}(H)$, it suffices to approximate the target function $f(x) = 1 / x$ on the interval $D_\kappa = [1/\kappa, 1]$ using an odd function. This allows us to focus on the desired range of the function and effectively approximate the matrix inversion operation.

\subsection{Constructing target polynomials approximating target functions}
To ensure numerical stability in the phase-factor evaluation method, we approximate the target functions using target polynomials that satisfy the conditions outlined in \cref{thm:sym_qsp}. Various methods have been proposed in the literature for constructing these polynomial approximations in a streamlined manner.

One approach is to directly truncate the Chebyshev series expansion of the target function. This can be efficiently achieved using Fast Fourier Transformation (FFT) applied to the transformed target function $f(\cos(\theta))$. However, when the target function is not defined on the entire interval $[-1, 1]$, the truncated series polynomial may not be bounded by $1$ on the entire interval, making it unsuitable for representation using QSP. To address this issue, one approach is to use the Remez exchange algorithm proposed in Ref.~\cite{DongMengWhaleyEtAl2021} to find the best polynomial approximation for the partially defined target function. Another method involves numerically finding the best polynomial approximation using a convex optimization-based approach as described in Ref.~\cite[Section IV]{DongLinTong2022}.

\begin{figure}[htbp]
        \centering
        \begin{subfigure}{0.45\textwidth}
                \centering
                \includegraphics[
                width =\textwidth
                ]{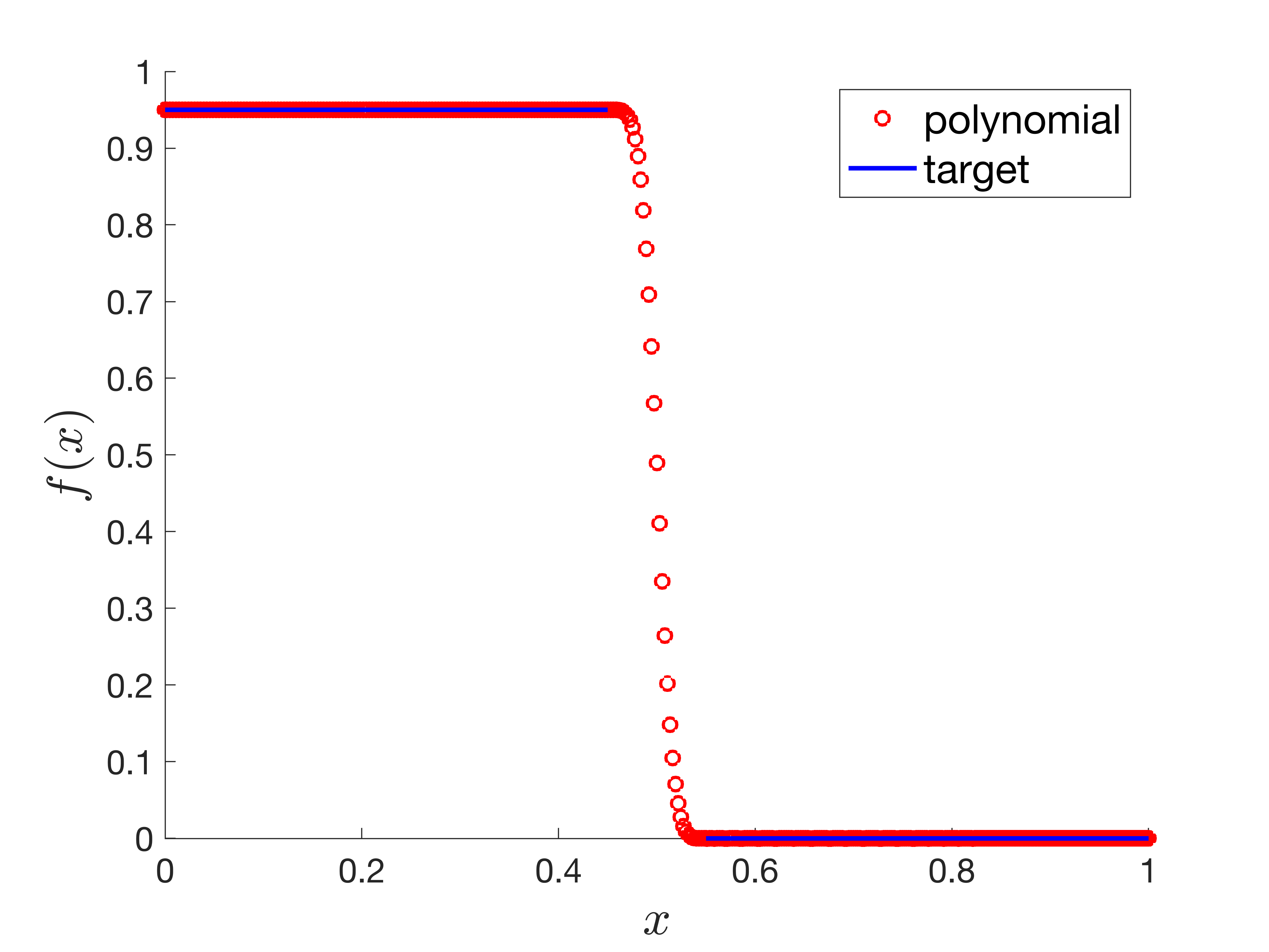}
                \caption{Heaviside energy filter function and its polynomial approximation with $\delta = 0.1$.}
        \end{subfigure}
        \hfill
        \begin{subfigure}{0.45\textwidth}
                \centering
                \includegraphics[
                width =\textwidth
                ]{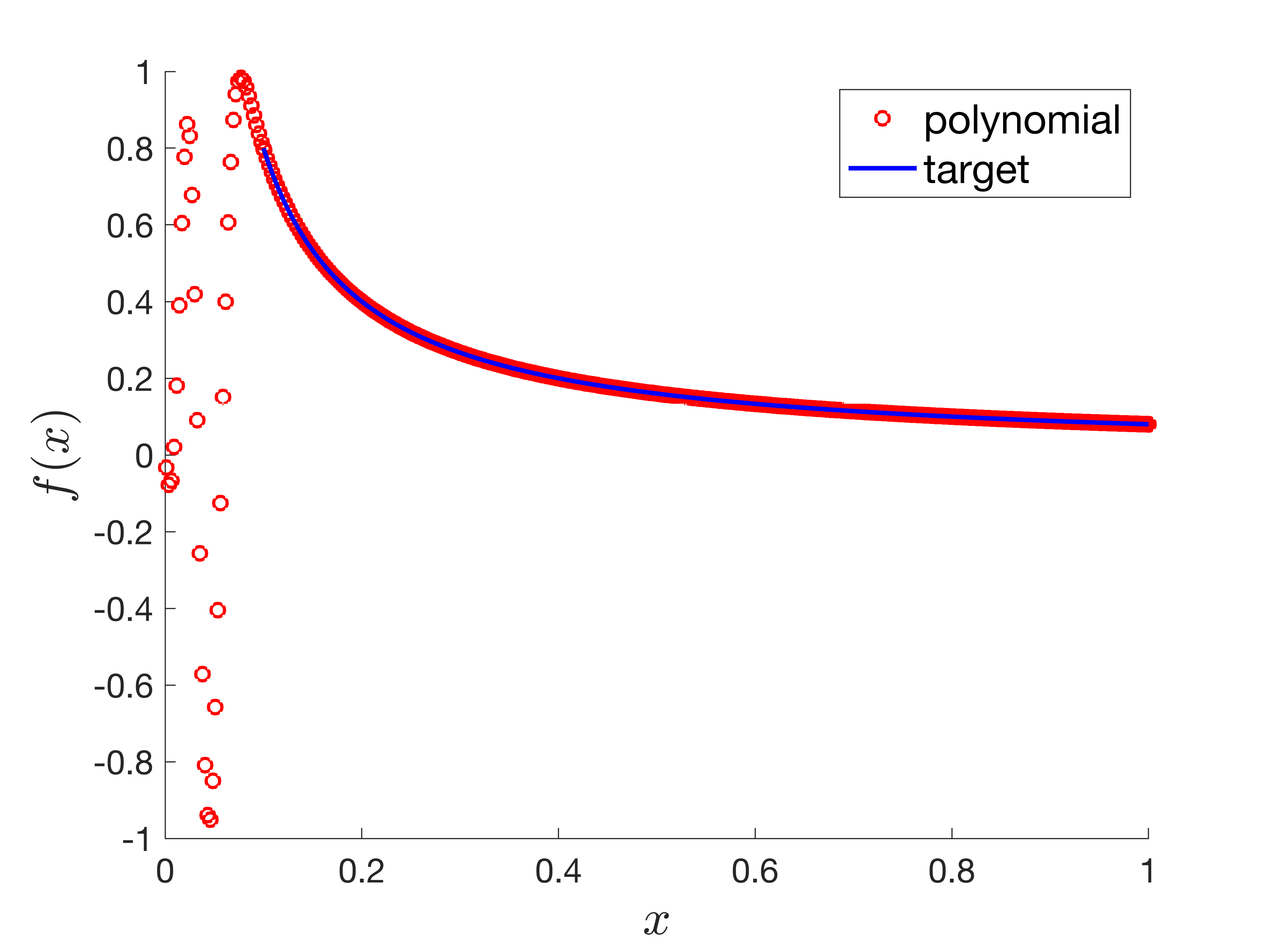}
                \caption{Matrix inversion function and its polynomial approximation with $\kappa = 10$.}
        \end{subfigure}
        \caption{Polynomial approximation of the target functions obtained by the convex-optimization-based method.\label{fig:polynomial_approximation}}
\end{figure}

In the presented numerical examples, we use the truncated Chebyshev series for quantum Hamiltonian simulation and quantum Gaussian filter, where the target functions are defined on the interval $[-1, 1]$. For other examples where the target function is defined on a further subinterval, we employ the convex optimization-based method to find the target polynomial approximation. The resulting target polynomials, obtained using the convex optimization-based method, are visualized in \cref{fig:polynomial_approximation}.

\subsection{Numerical results}
We evaluate the performance of Newton's method for finding phase factors in the presented numerical tests. All experiments are conducted using \textsf{Matlab R2020a} on a computer with an Intel Core i5 Quad CPU running at 2.11 GHz and 8 GB of RAM.

\begin{figure}[htbp]
        \centering
        \begin{subfigure}{0.45\textwidth}
                \centering
                \includegraphics[
                width =\textwidth
                ]{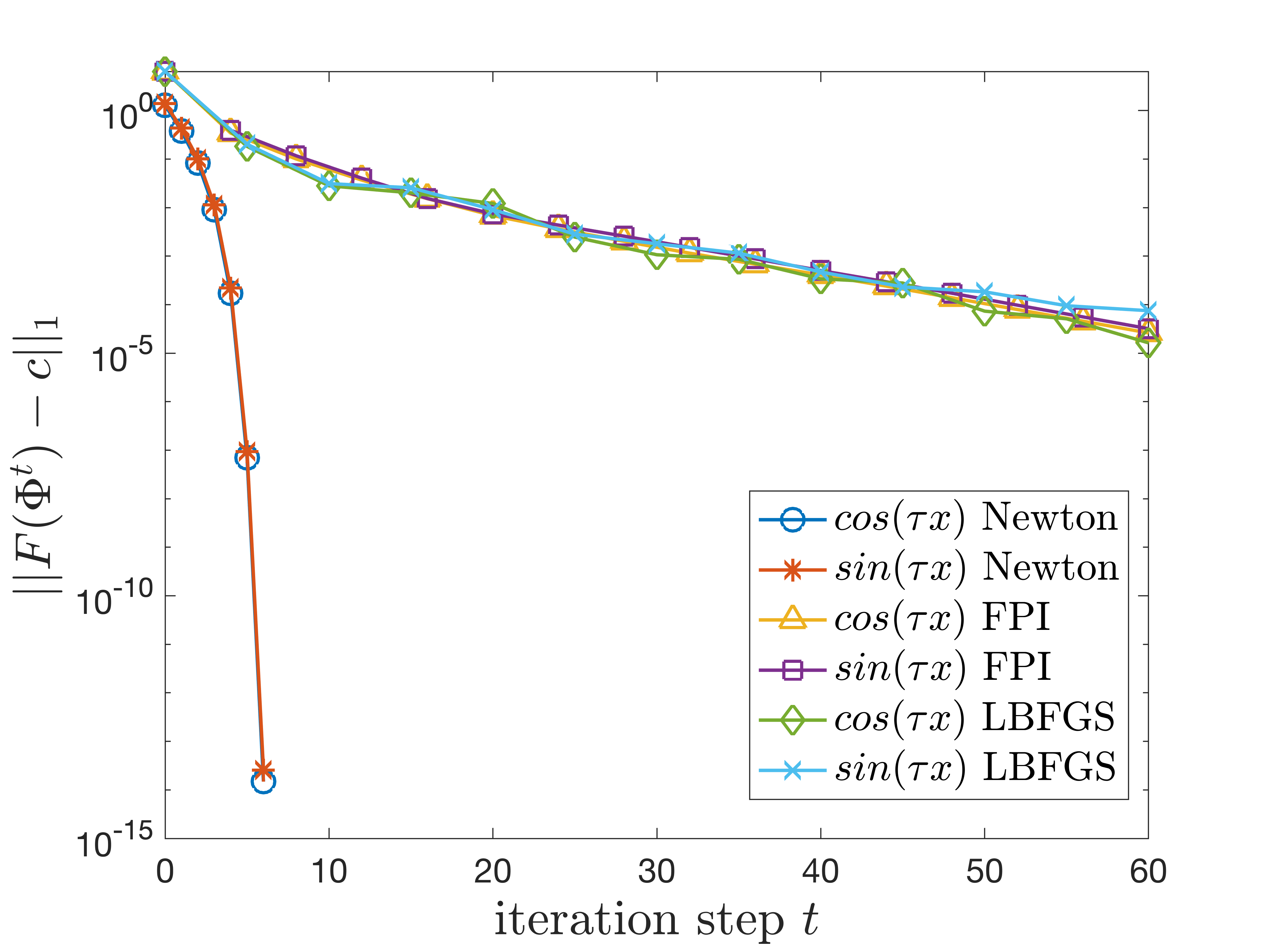}
                \caption{ Hamiltonian simulation}
        \end{subfigure}
        \hfill
        \begin{subfigure}{0.45\textwidth}
                \centering
                \includegraphics[
                width =\textwidth
                ]{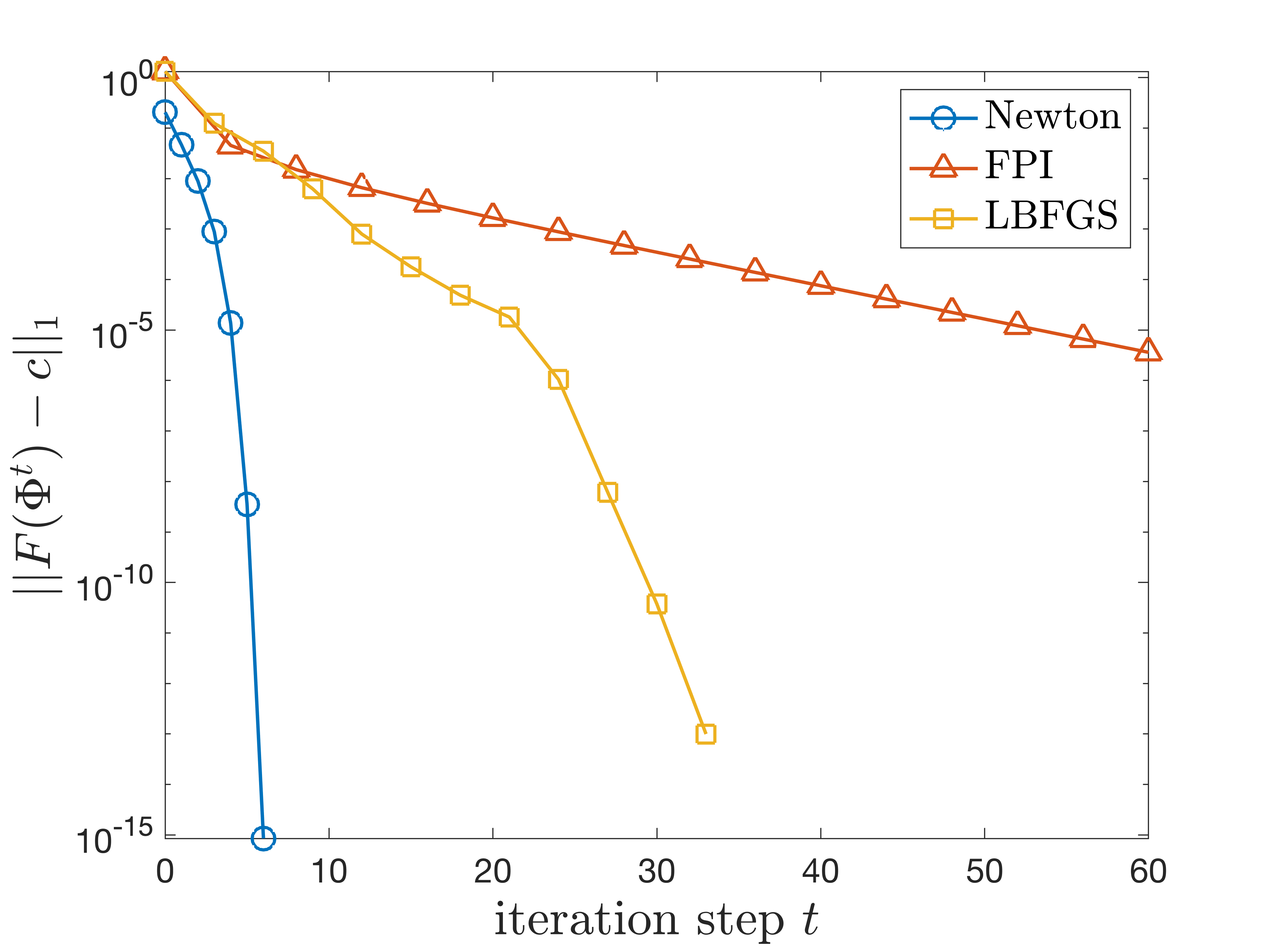}
                \caption{ Gaussian filter }
        \end{subfigure}
        \hfill
        \begin{subfigure}{0.45\textwidth}
                \centering
                \includegraphics[
                width =\textwidth
                ]{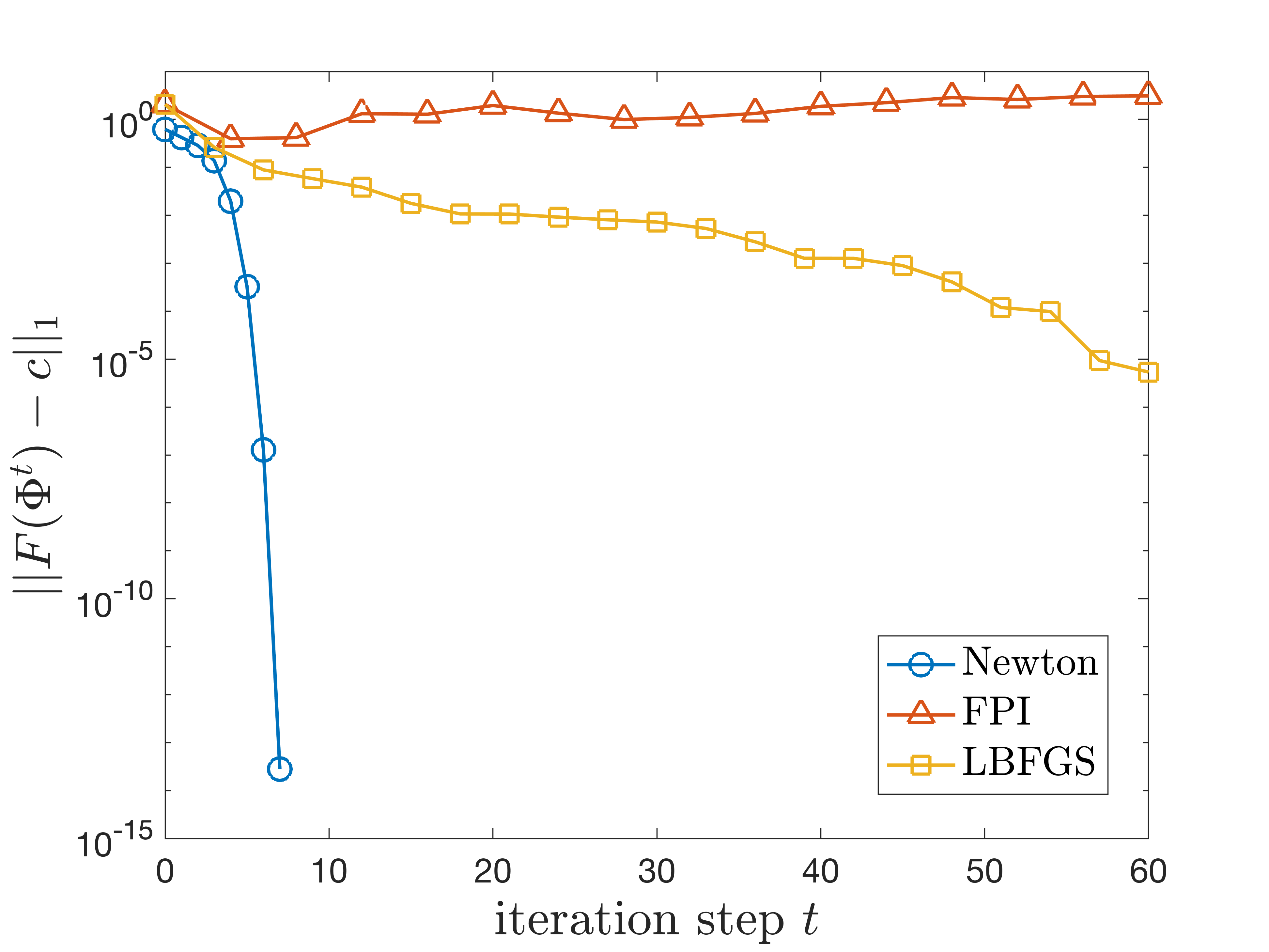}
                \caption{Heaviside energy filter}
        \end{subfigure}
        \hfill
        \begin{subfigure}{0.45\textwidth}
                \centering
                \includegraphics[
                width =\textwidth
                ]{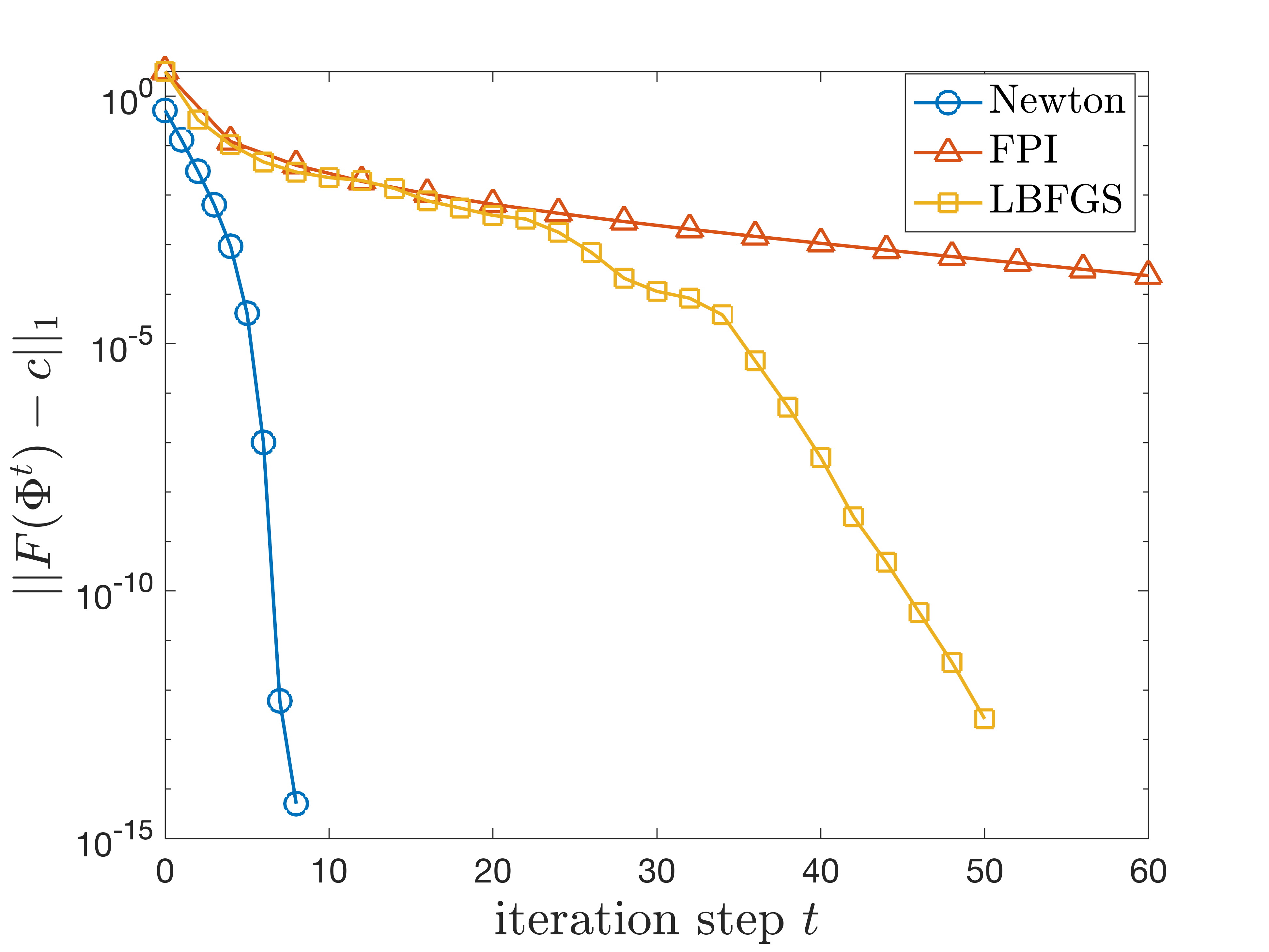}
                \caption{Matrix inversion}
        \end{subfigure}
        \caption{Residual error as a function of iteration steps for different numerical examples. The target polynomials in (a)-(d) are chosen to be near the fully-coherent regime, with the maximum absolute value set to 0.99 in (a)-(c) and 0.998 in (d). (a) Quantum Hamiltonian simulation with $\tau = 100$. The target polynomials, obtained by truncating the Jacobi-Anger series with truncation error $\epsilon_0 = 10^{-14}$, have degrees of 1390 and 1391. (b) Quantum Gaussian filter with $\mu = 0.5$ and $\sigma = 0.1$. The target polynomial is derived from the Chebyshev series expansion using FFT, resulting in a degree-100 polynomial. (c) Heaviside energy filter with $\delta = 0.1$. The target polynomial is a degree-250 polynomial obtained from a convex-optimization-based method. (d) Matrix inversion with $\kappa = 10$. The target polynomial is a degree-301 polynomial derived from a convex-optimization-based method.\label{fig:numerical_error}}
\end{figure}

\begin{figure}[htbp]
        \centering
        \begin{subfigure}{0.45\textwidth}
                \centering
                \includegraphics[
                width =\textwidth
                ]{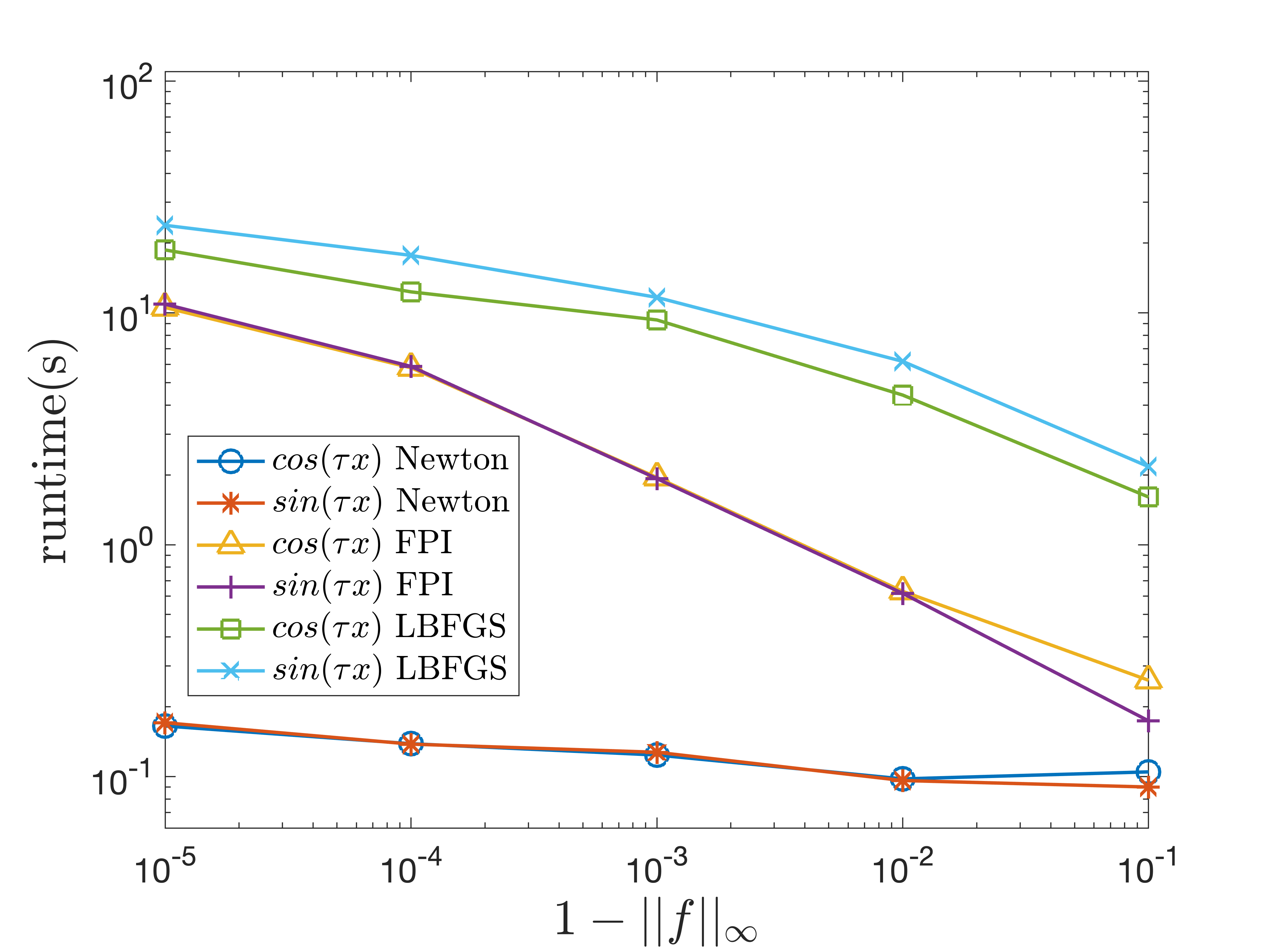}
                \caption{Hamiltonian simulation}
        \end{subfigure}
        \hfill
        \begin{subfigure}{0.45\textwidth}
                \centering
                \includegraphics[
                width =\textwidth
                ]{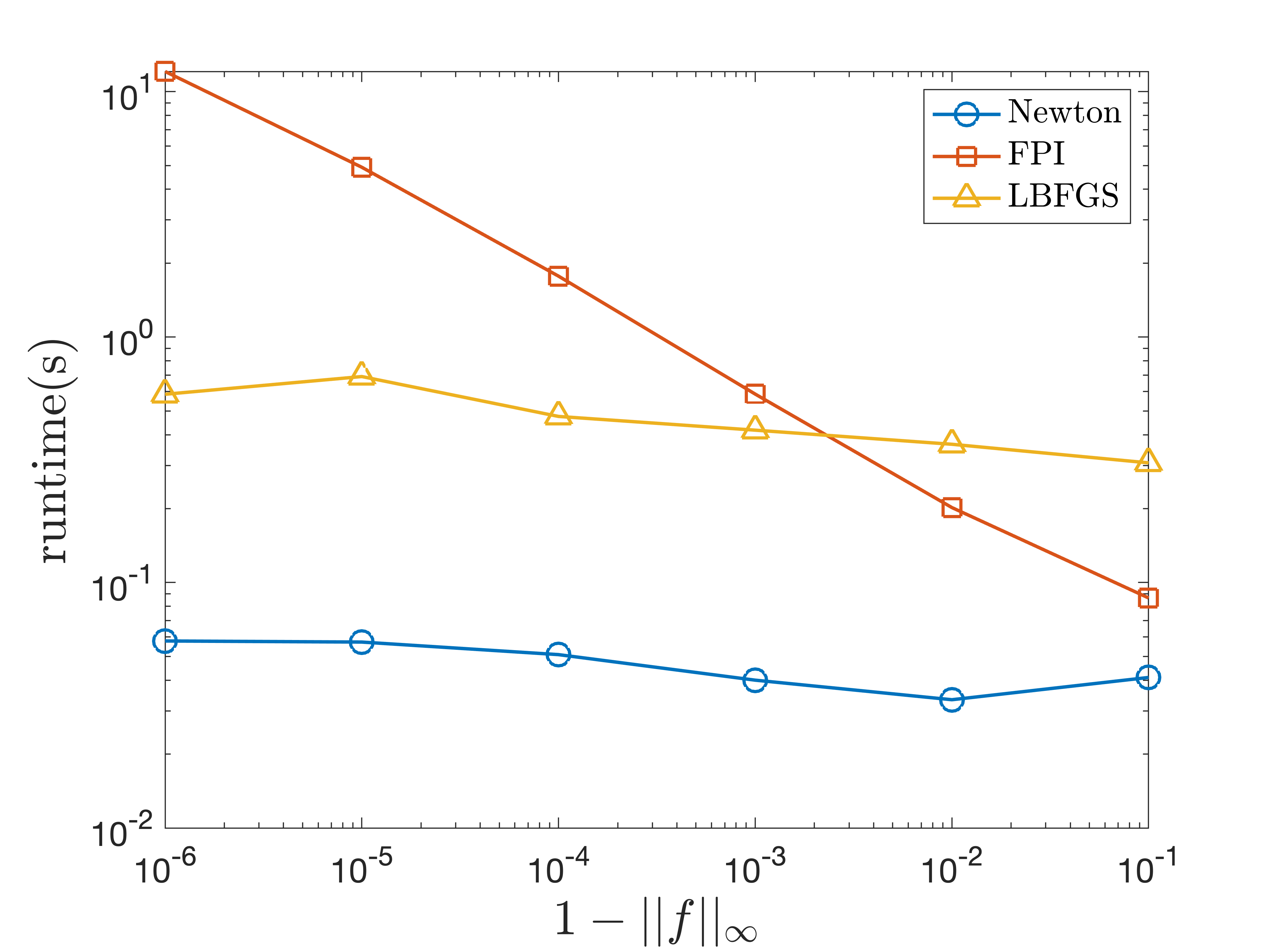}
                \caption{Gaussian filter}
        \end{subfigure}
        \caption{Runtime analysis of numerical examples for different target functions. The target polynomials in (a) and (b) are scaled by a constant to make the problem increasingly close to the fully-coherent regime $\norm{f}_\infty \to 1$. (a) Quantum Hamiltonian simulation with $\tau = 100$ and truncation error $\epsilon_0 = 1 \times 10^{-14}$. The target polynomials have degrees of 167 and 168. (b) Quantum Gaussian filter with $\mu = 0.5$ and $\sigma = 0.01$. The target polynomial is of degree 100.\label{fig:numerical_runtime}}
\end{figure}

The performance metrics used to evaluate the performance of Newton's method are the runtime and the residual error. The runtime refers to the amount of time it takes for the method to converge and find the desired phase factors. The residual error measures the discrepancy between the polynomial parametrized by the computed phase factors and the true polynomial which is defined as
\begin{equation}
    \mathrm{residual\ error} = \norm{F(\Phi)-c}_1.
\end{equation}
The numerical results for the error metric of Newton's method are presented in \cref{fig:numerical_error}. It is evident from the results that Newton's method exhibits significantly faster convergence compared to other iterative methods for solving phase factors. The error curve aligns well with the expected quadratic convergence of Newton's method in general analysis. Besides, Newton's method exhibits greater stability in terms of runtime as the target function approaches the fully-coherent regime. \cref{fig:numerical_runtime} depicts the numerical results for the runtime of three iterative methods for determining phase factors. It also clearly illustrates the superior speed of Newton's method compared to the other two iterative methods.

\begin{figure}[htbp]
        \centering
        \begin{subfigure}{0.45\textwidth}
                \centering
                \includegraphics[
                width =\textwidth
                ]{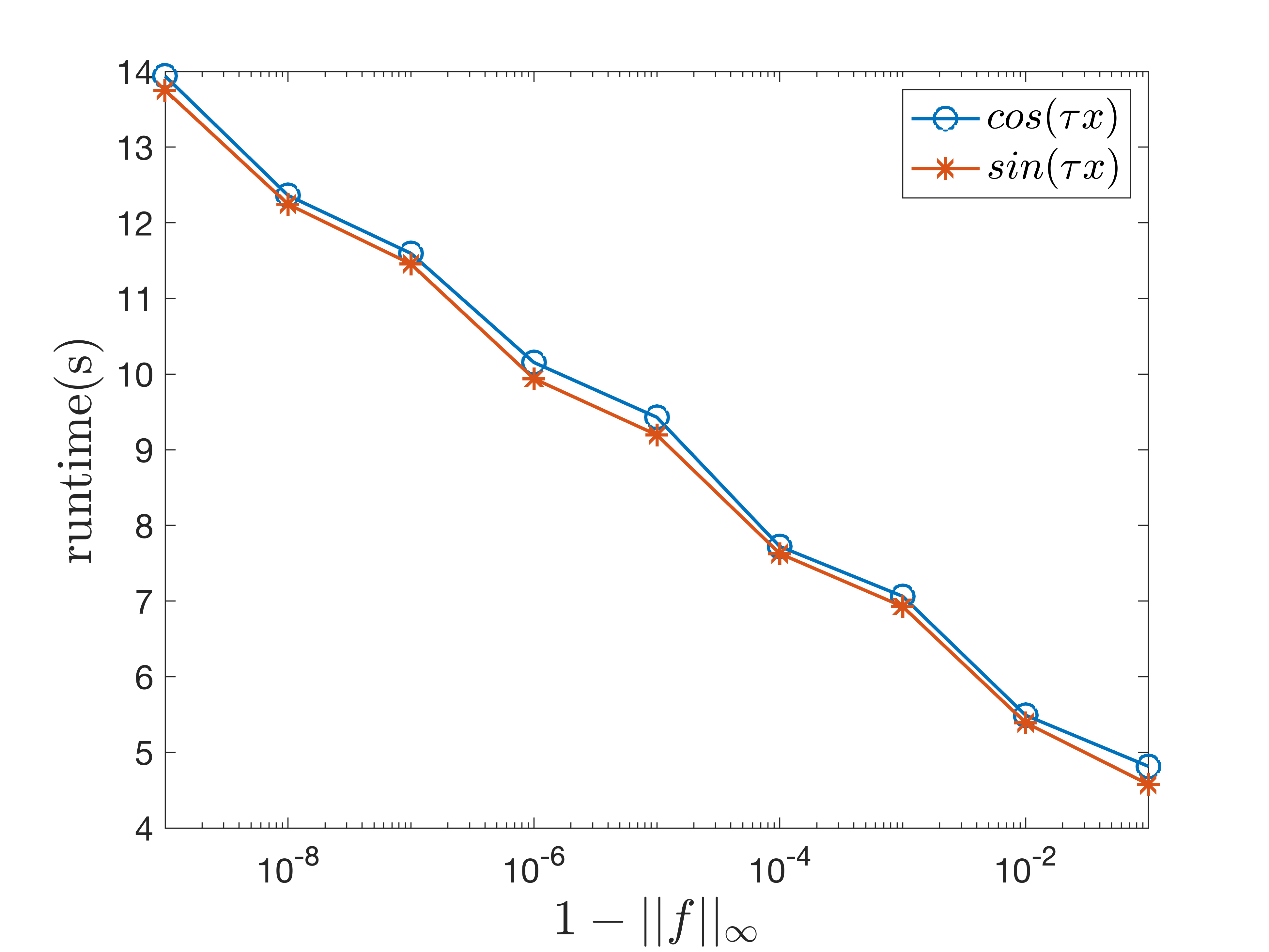}
                \caption{}
        \end{subfigure}
        \hfill
        \begin{subfigure}{0.45\textwidth}
                \centering
                \includegraphics[
                width =\textwidth
                ]{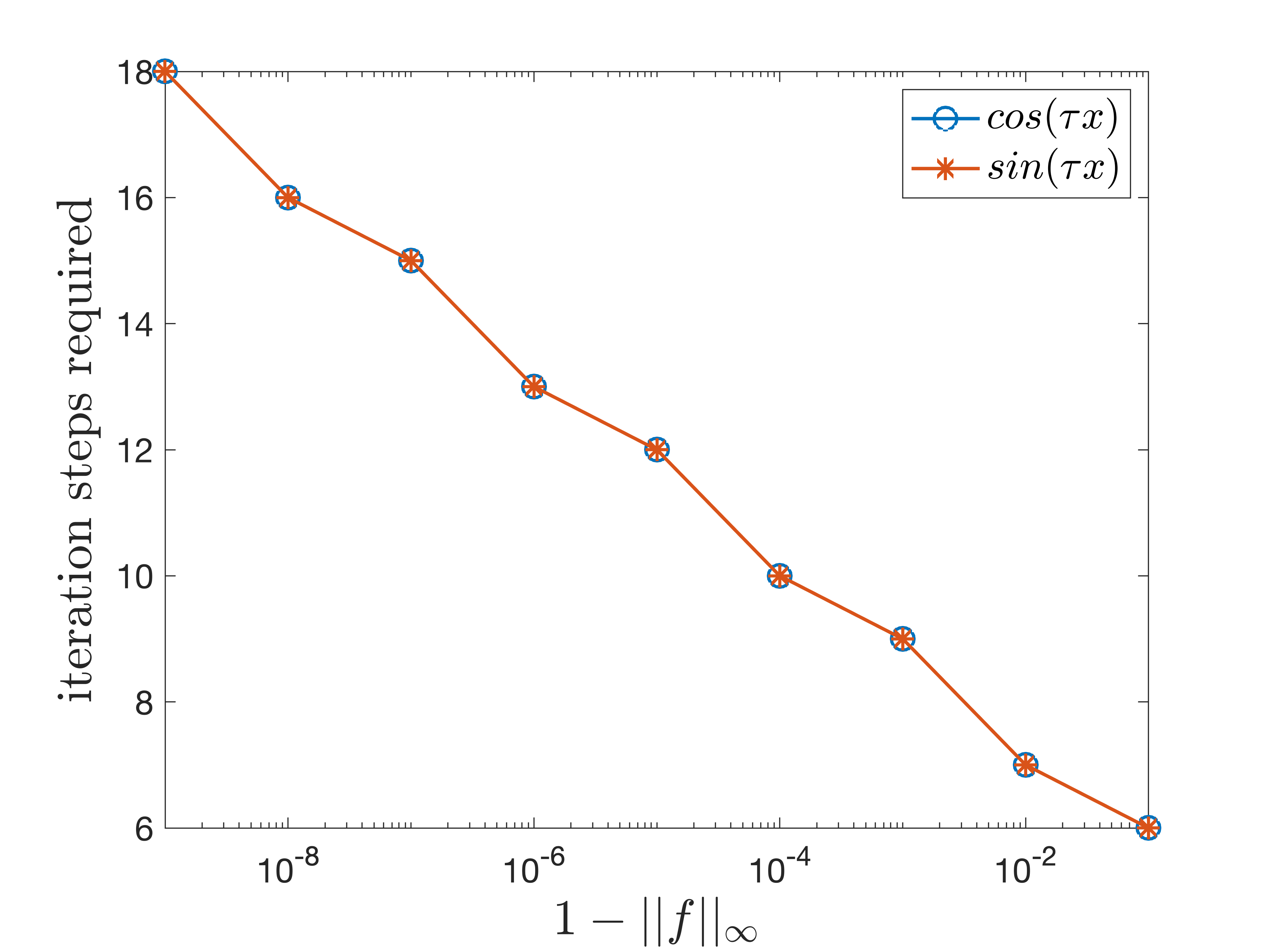}
                \caption{}
        \end{subfigure}
        \caption{Convergence analysis of Newton's method near the fully-coherent regime. The target polynomial in each example is scaled by a constant to make the problem increasingly close to the fully-coherent regime $\norm{f}_\infty \to 1$. The problem is set to be quantum Hamiltonian simulation with $\tau = 1000$ and truncation error $\epsilon_0 = 1 \times 10^{-14}$. The degrees of the target polynomials are 1390 and 1391. (a) Runtime of Newton's method. (b) Number of iterations before convergence.\label{fig:runtime_to_fully_coherent}}
\end{figure}

To further analyze the performance of Newton's method near the fully-coherent regime, the runtime and the number of iterations are plotted as a function of the distance to the fully-coherent regime ($1 - \norm{f}_\infty$) in \cref{fig:runtime_to_fully_coherent}. It is noteworthy that even when the target function is extremely close to being fully-coherent ($1 - \norm{f}_\infty \le 1 \times 10^{-9}$), Newton's method is capable of locating the optimum within a small number of iterations. This result highlights the robustness of Newton's method for finding phase factors in the nearly fully-coherent regime.

\begin{figure}[htbp]
      \centering
      \begin{subfigure}{0.45\textwidth}
              \centering
              \includegraphics[
              width =\textwidth
              ]{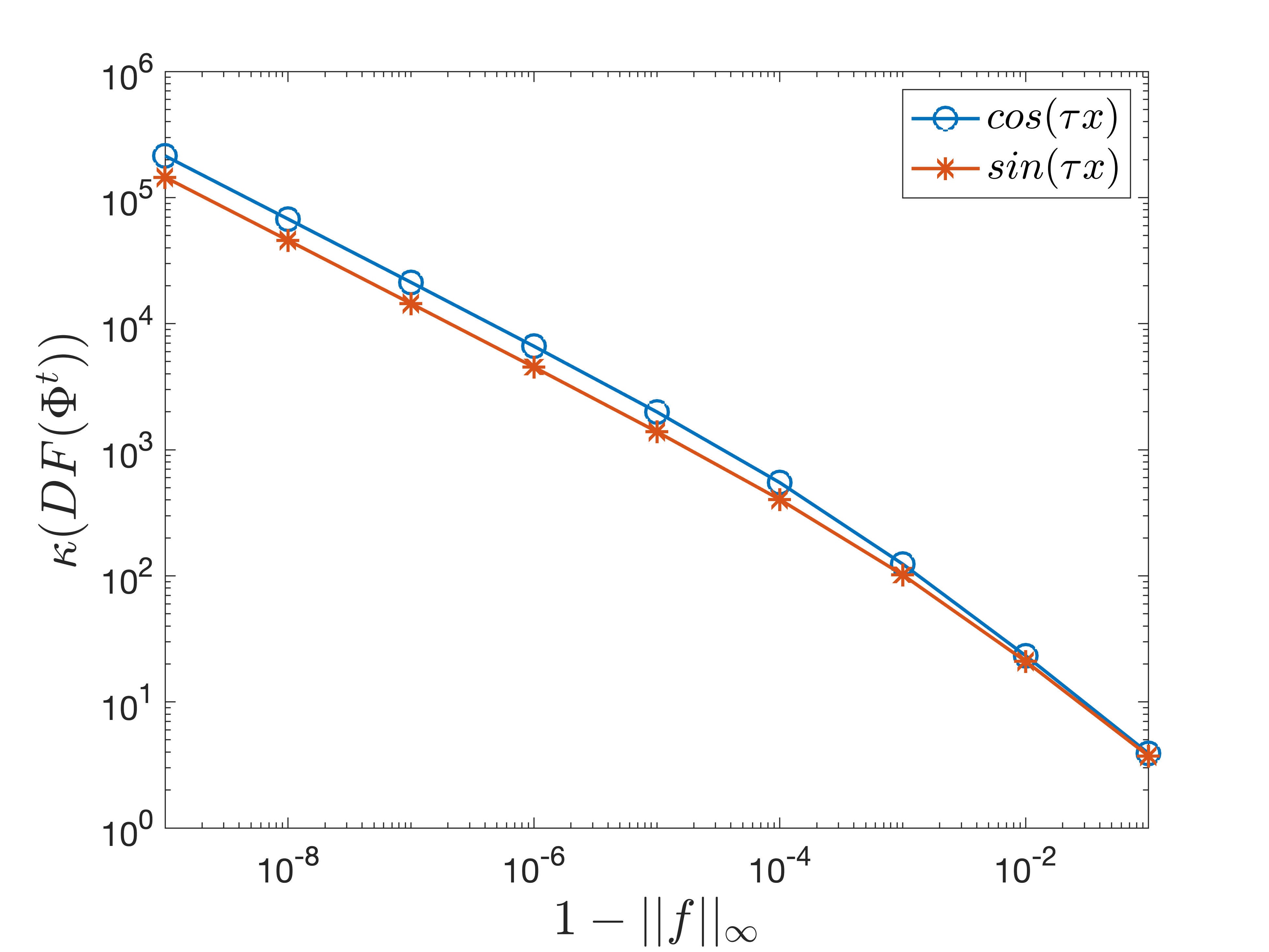}
              \caption{Hamiltonian simulation}
      \end{subfigure}
      \hfill
      \begin{subfigure}{0.45\textwidth}
              \centering
              \includegraphics[
              width =\textwidth
              ]{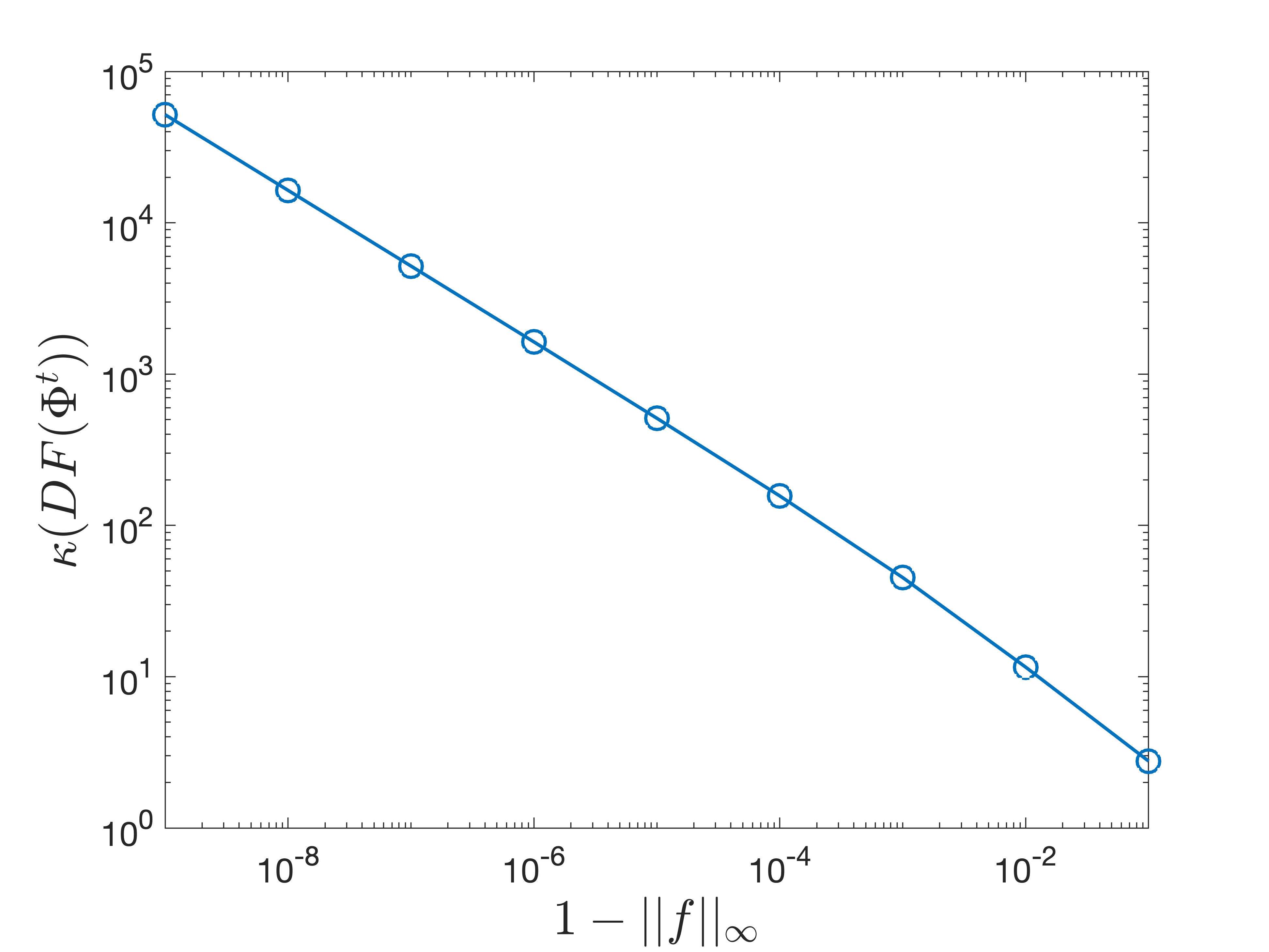}
              \caption{Gaussian filter}
      \end{subfigure}
      \caption{Condition number of the Jacobian matrix at the numerical optimum. Each problem is solved by using Newton's method. The target polynomial in each example is scaled by a constant to make the problem increasingly close to the fully-coherent regime. (a) Quantum Hamiltonian simulation with $\tau = 1000$ and truncation error $\epsilon_0 = 1 \times 10^{-14}$. The target polynomials have degrees of 1390 and 1391. (b) Quantum Gaussian filter with $\mu = 0.5$ and $\sigma = 0.01$. The target polynomial is of degree 100.\label{fig:Jacobian_cond_fully_coherent}}
\end{figure}

Finally, we investigate the condition number of the Jacobian matrices at the phase factors obtained by Newton's method for different target functions, as presented in \cref{fig:Jacobian_cond_fully_coherent}. The results indicate that as the target function approaches the fully-coherent regime, the condition number of the Jacobian matrix becomes increasingly ill-conditioned. Despite this challenge, Newton's method continues to exhibit remarkable performance in finding phase factors. This emphasizes the effectiveness and reliability of Newton's method in phase factor determination, even in challenging scenarios near the fully-coherent regime.

\section{Conclusion}

This paper presents a novel approach to solving the vector-valued, nonlinear system that arises in quantum signal processing (QSP) using Newton's method. Numerical results indicate that the proposed method can robustly find phase factors in all parameter regimes, in particular the challenging fully-coherent regime with ill-conditioned Jacobian matrices. Our method takes advantage of the matrix product states structure of QSP, enabling efficient computation of the Jacobian matrix. Additionally, the use of real-number arithmetics further enhances the prefactor of the numerical method. The method has been implemented in the QSPPACK software package, providing a practical tool for solving QSP problems in scientific computing on quantum computers.

From a theoretical perspective, there are open problems regarding the impressive performance of Newton's method. While convergence in an $\ell^1$ neighborhood of $\mathbf{0}$ can be understood using the same contraction mapping technique as in~\cite{DongLinNiEtAl2022}, the theoretical understanding of the effectiveness of the method in the fully-coherent regime remains a mystery. Additionally, extensive numerical experiments consistently converged to the maximal solution which is a special class of symmetric phase-factor solutions proposed and studied in Ref.~\cite{WangDongLin2022}. Further investigations are needed to understand whether the mapping $F$ admits a unique landscape within the injective neighborhood near $\mathbf{0}$.

\textbf{Acknowledgment}

This material is based upon work supported by the U.S. Department of Energy, Office of Science, National Quantum Information Science Research Centers, Quantum Systems Accelerator (J.W.). Additional support is acknowledged from NSF Quantum Leap Challenge Institute (QLCI) program under Grant number OMA-2016245 (Y.D.), the Applied Mathematics Program of the US Department of Energy (DOE) Office of Advanced Scientific Computing Research under contract number DE-AC02-05CH1123, and the Google Quantum Research Award. L.L. is a Simons Investigator.

\appendix

\section{Details about the formalism of symmetric QSP in real arithmetic operations}\label{sec:appendix_QSP_real_arithmetic}
In the main text, we present a concise idea of the real-number arithmetic representation of QSP. In this section, we aim to provide a more comprehensive discussion and present additional details on this topic.

The computation of the QSP matrix boils down to that of a sequence of unitary matrix multiplications in \cref{eq:UPQ}. Furthermore, the QSP matrix admits the following decomposition as a consequence of \cref{thm:sym_qsp}
\begin{equation}\label{eqn:QSP_decomposition}
U(x,\Phi)= \begin{pmatrix}
a_{\wt{d}-1}(x) + \I d_{\wt{d}-1}(x) & \I \alpha_{\wt{d}-1} (x)\\
\I \alpha_{\wt{d}-1} (x) & a_{\wt{d}-1}(x) - \I d_{\wt{d}-1} (x)
\end{pmatrix}.
\end{equation}
Here, $a_{\wt{d}-1}(x)$, $d_{\wt{d}-1}(x)$ and $\alpha_{\wt{d}-1}(x)/\sqrt{1-x^2}$ are real polynomials in the variable $x$. According to the convention presented in the main text, $d_{\wt{d}-1}$ stands for the component of interest, also known as $g(x,\Phi)$ in the main text to emphasize the dependence in phase factors $\Phi$. As the goal in this section is to derive a simple recipe for computing the QSP matrix with a given set of phase factors $\Phi$, we drop the $\Phi$ dependence in this section for the notational simplicity.

Let the entry-wise value of the phase factors be $\Phi=(\phi_0, \phi_1,\cdots,\phi_{\wt{d}-1})$. For ease of discussion, we refer to $\Phi^{(k)} = (\phi_0, \phi_1, \cdots, \phi_k)$ as the $k$-th truncated phase factors for each $k = 0, 1,\cdots, \wt{d}-1$. The corresponding sequence of QSP matrices is denoted entry-wise as
\begin{equation}
    U(x,\Phi^{(k)})= \begin{pmatrix}
a_k(x) + \I d_k(x) & \I \alpha_k (x)\\
\I \alpha_k (x) & a_k(x) - \I d_k (x)
\end{pmatrix}.
\end{equation}
We remark that each truncated set of phase factors also gives a symmetric QSP. Hence, the decomposition in \cref{eqn:QSP_decomposition} applies, implying that $a_k(x)$, $\alpha_k(x)$ and $d_k(x)$ are well defined. By appending $\phi_{k}$ to the $(k-1)$-th truncation $\Phi^{(k-1)}$, the recurrence relation follows
\begin{equation}\label{eqn:complex_recurrence}
\begin{split}
   &\begin{pmatrix}
a_k(x) + \I d_k(x) & \I \alpha_k (x)\\
\I \alpha_k (x) & a_k(x) - \I d_k (x)
\end{pmatrix}\\
&=e^{\I \phi_k Z} W(x) \begin{pmatrix}
a_{k-1}(x) + \I d_{k-1}(x) & \I \alpha_{k-1} (x)\\
\I \alpha_{k-1} (x) & a_{k-1}(x) - \I d_{k-1} (x)
\end{pmatrix}  W(x) e^{\I \phi_k Z}.
\end{split}
\end{equation}
It can be verified that the following rearrangement is equivalent to the recurrence relation
\begin{equation}\label{eq:induction}
    \begin{pmatrix}
    a_{k} (x)\\
    d_{k} (x) \\
    \alpha_{k} (x)
    \end{pmatrix} = R_z(2\phi_k) R_x(2\arccos(x)) \begin{pmatrix}
    a_{k-1} (x)\\
    d_{k-1} (x) \\
    \alpha_{k-1} (x)
    \end{pmatrix},
\end{equation}
where 
\begin{equation}
\begin{split}
    & R_z(2 \phi) = \begin{pmatrix}
        \cos{2\phi} & -\sin{2\phi} & \\
        \sin{2\phi} & \cos{2\phi} & \\
        & & 1
        \end{pmatrix}\\
    \text{ and } &
    R_x(2\arccos(x)) = \begin{pmatrix}
    2x^2-1 & & -2x\sqrt{1-x^2}\\
     & 1 & \\
     2x\sqrt{1-x^2} & & 2x^2-1
    \end{pmatrix}
\end{split}
\end{equation}
are the induced $\mathrm{SO}(3)$ rotation matrices. It can also be shown that the base cases of the recurrence are

(1) when $d$ is even
\begin{equation}\label{eqn:base-even}
    \begin{pmatrix}
    a_0(x) \\
    d_0(x)\\
    \alpha_0(x)
    \end{pmatrix}=
        \begin{pmatrix}
    \cos(2\phi_0)\\
    \sin(2\phi_0)\\
    0
    \end{pmatrix},
\end{equation}
and (2) when $d$ is odd
\begin{equation}\label{eqn:base:odd}
    \begin{pmatrix}
    a_0(x) \\
    d_0(x)\\
    \alpha_0(x)
    \end{pmatrix}= \begin{pmatrix}
    \cos(2\phi_0) x\\
    \sin(2\phi_0) x\\
    \sqrt{1-x^2}
    \end{pmatrix}.
\end{equation}
Remarkably, the equivalent recurrence relation \cref{eq:induction} involves only real quantities, and $R_x$ and $R_z$ only actively act as a rotation on two entries. In contrast to the complex recurrence relation \cref{eqn:complex_recurrence}, the real recurrence has lower time and space complexity. This improvement is due to the simplified structure of symmetric QSP compared with the original formalism without symmetry.

The MPS/TT structure still holds in the real recurrence relation. We refer $\mc{X}$ and $\mc{Z}$ to the parametric order-$2$ tensor standing for the induced $\mathrm{SO}(3)$ rotations, namely, $\mc{Z}(\phi) = R_z(2\phi)$ and $\mc{X}(x) = R_x(2\arccos(x))$. Let $\mc{I}$ be the order-$1$ tensor representing the base of the recurrence in \cref{eqn:base-even,eqn:base:odd}. Furthermore, to extract the component of the computational interest, let $\mc{H}$ be the order-$1$ tensor representing the last operation, which is
\begin{equation}
    \mc{H}(\phi_{\wt{d}-1}) = \begin{pmatrix}
        0\ 1\ 0
    \end{pmatrix} R_z(2\phi_{\wt{d}-1}) = \begin{pmatrix}
        \sin(2\phi_{\wt{d}-1}) \ \cos(2\phi_{\wt{d}-1}) \ 0
    \end{pmatrix}.
\end{equation}

\begin{figure}[tbhp]
\centering
\begin{subfigure}{\textwidth}
\centering
\scalebox{1}{\begin{tikzpicture}
    \node[draw, shape=circle] (v0) at (0,0) {$\mc{H}$};
    \node[draw, shape=circle] (v1) at (1,0) {$\mc{X}$};
    \node[draw, shape=circle] (v2) at (2,0) {$\mc{Z}$};
    \node(v3) at (3,0) {$\cdots$};
    \node[draw, shape=circle] (v4) at (4,0) {$\mc{X}$};
    \node[draw, shape=circle] (v5) at (5,0) {$\mc{Z}$};
    \node[draw, shape=circle] (v6) at (6,0) {$\mc{X}$};
    \node(v7) at (7,0) {$\cdots$};
    \node[draw, shape=circle] (v8) at (8,0) {$\mc{Z}$};
    \node[draw, shape=circle] (v9) at (9,0) {$\mc{X}$};
    \node[draw, shape=circle] (v10) at (10,0) {$\mc{I}$};
    \node (u0) at (0,-1) {$\phi_{\wt{d}-1}$};
    \node (u1) at (1,-1) {$x$};
    \node (u2) at (2,-1) {$\phi_{\wt{d}-2}$};
    \node (u4) at (4,-1) {$x$};
    \node (u5) at (5,-1) {$\phi_i$};
    \node (u6) at (6,-1) {$x$};
    \node (u8) at (8,-1) {$\phi_1$};
    \node (u9) at (9,-1) {$x$};
    \node (u10) at (10,-1) {$\phi_0$};
    \draw [thick] (v0) -- (v1) -- (v2) -- (v3) -- (v4) -- (v5) -- (v6) -- (v7) -- (v8) -- (v9) -- (v10)
    (v0) -- (u0)
    (v1) -- (u1)
    (v2) -- (u2)
    (v4) -- (u4)
    (v5) -- (u5)
    (v6) -- (u6)
    (v8) -- (u8)
    (v9) -- (u9)
    (v10) -- (u10);
\end{tikzpicture}}
\caption{}
\end{subfigure}
\hfill
\begin{subfigure}{\textwidth}
\centering
    \scalebox{1}{\begin{tikzpicture}[decoration={brace,mirror,amplitude=7}]
    \node[draw, shape=circle] (v0) at (0,0) {$\mc{H}$};
    \node[draw, shape=circle] (v1) at (1,0) {$\mc{X}$};
    \node[draw, shape=circle] (v2) at (2,0) {$\mc{Z}$};
    \node(v3) at (3,0) {$\cdots$};
    \node[draw, shape=circle] (v4) at (4,0) {$\mc{X}$};
    \node[draw, shape=circle] (v5) at (5,0) {$\mc{Z}$};
    \node[draw, shape=circle] (v6) at (6,0) {$\mc{X}$};
    \node(v7) at (7,0) {$\cdots$};
    \node[draw, shape=circle] (v8) at (8,0) {$\mc{X}$};
    \node[draw, shape=circle] (v9) at (9,0) {$\mc{I}$};
    \node (u0) at (0,-1) {$\phi_{\wt{d}-1}$};
    \node (u1) at (1,-1) {$x$};
    \node (u2) at (2,-1) {$\phi_{\wt{d}-2}$};
    \node (u4) at (4,-1) {$x$};
    \node (u5) at (5,-1) {$\phi_i$};
    \node (u6) at (6,-1) {$x$};
    \node (u8) at (8,-1) {$x$};
    \node (u9) at (9,-1) {$\phi_{0}$};
    \node (e1) at (4.5,0.6) {};
    \node (e2) at (4.5,-0.6) {};
    \node (e3) at (5.5,0.6) {};
    \node (e4) at (5.5,-0.6) {};
    \draw [thick] (v0) -- (v1) -- (v2) -- (v3) -- (v4) -- (v5) -- (v6) -- (v7) -- (v8) -- (v9) 
    (v0) -- (u0)
    (v1) -- (u1)
    (v2) -- (u2)
    (v4) -- (u4)
    (v5) -- (u5)
    (v6) -- (u6)
    (v8) -- (u8)
    (v9) -- (u9);
    \draw[dashed] (e1) -- (e2) (e3)--(e4);
    \draw [decorate] ([yshift=-3mm]u0.west) --node[below=3mm]{$\mc{N}_\text{left}^{(i)}$} ([yshift=-3mm]u4.east);
    \draw [decorate] ([yshift=-3mm]u6.west) --node[below=3mm]{$\mc{N}_\text{right}^{(i)}$} ([yshift=-3mm]u9.east);
\end{tikzpicture}}
\caption{}
\end{subfigure}
\caption{A graphical visualization of the MPS/TT structure of the problem in the real-number arithmetic representation. (a) The structure of the recurrence relation. (b) The partition when computing the Jacobian.}
\label{fig:real_visual}
\end{figure}
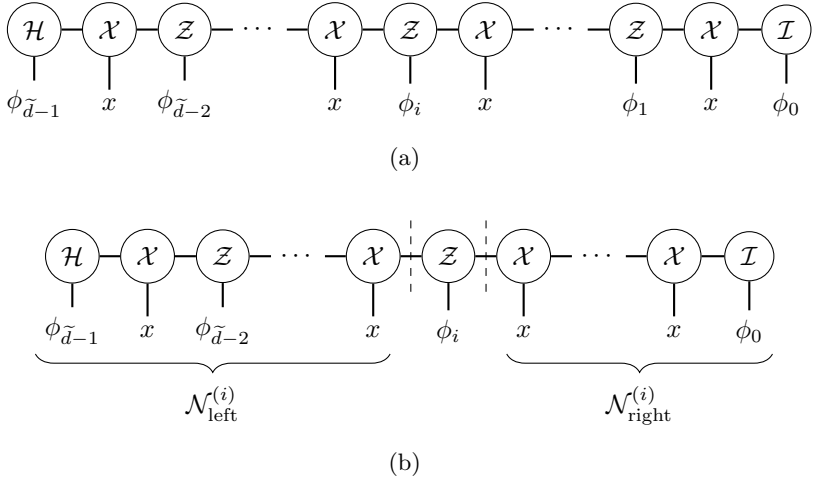

Then, the recurrence relation in real-number arithmetic can be visualized graphically in \cref{fig:real_visual}.
In contrast to the computation in the complex-arithmetic representation, the symmetry constraint of the QSP phase factors is reflected in the doubled argument in the $Z$ tensor of the real-number arithmetic representation. Hence, when computing the derivative, it does not need tricks to arrange the derivatives coming from two symmetric sites. Specifically, the following identity holds
\begin{equation}
    \frac{\partial g(x, \Phi)}{\partial \phi_i} = \mc{N}_\mathrm{left}^{(i)} \frac{\ud \mc{Z}(\phi_i)}{\ud \phi_i} \mc{N}_\mathrm{right}^{(i)} = 2 \mc{N}_\mathrm{left}^{(i)} \mc{Z}(\phi_i + \pi / 4) \mc{N}_\mathrm{right}^{(i)}.
\end{equation}
Here, the left and right parts under the partition are given by
\begin{equation}
    \begin{split}
        \mc{N}_{\text{left}}^{(i)}:= \mc{H}\prod_{j=\wt{d}-2}^{i+1}  \left(\mc{X}(x) \mc{Z}(\phi_j) \right)\mc{X}(x),\quad \mc{N}^{(i)}_{\text{right}}:= \prod_{j=i-1}^1 \mc{X}(x) \mc{Z}(\phi_j) \mc{I},
    \end{split}
\end{equation}
whose graphical visualizations are presented in \cref{fig:real_visual}. The update of these quantities in the computational process is
\begin{equation}
    \mc{N}_{\text{left}}^{(i+1)} \leftarrow \mc{N}_{\text{left}}^{(i)}\mc{X}^{-1}(x) \mc{Z}(-\phi_{i+1}),\quad\text{and}\quad\mc{N}^{(i+1)}_{\text{right}} \leftarrow \mc{Z}(\phi_i)\mc{X}(x) \mc{N}^{(i)}_{\text{right}}.
\end{equation}

For completeness, we provide the algorithm for computing the Jacobian matrix using the MPS/TT structure and the real-number arithmetic representation in \cref{alg:DF_real}.

\begin{figure}[htbp]
        \centering
        \includegraphics[
        width=0.7\textwidth]{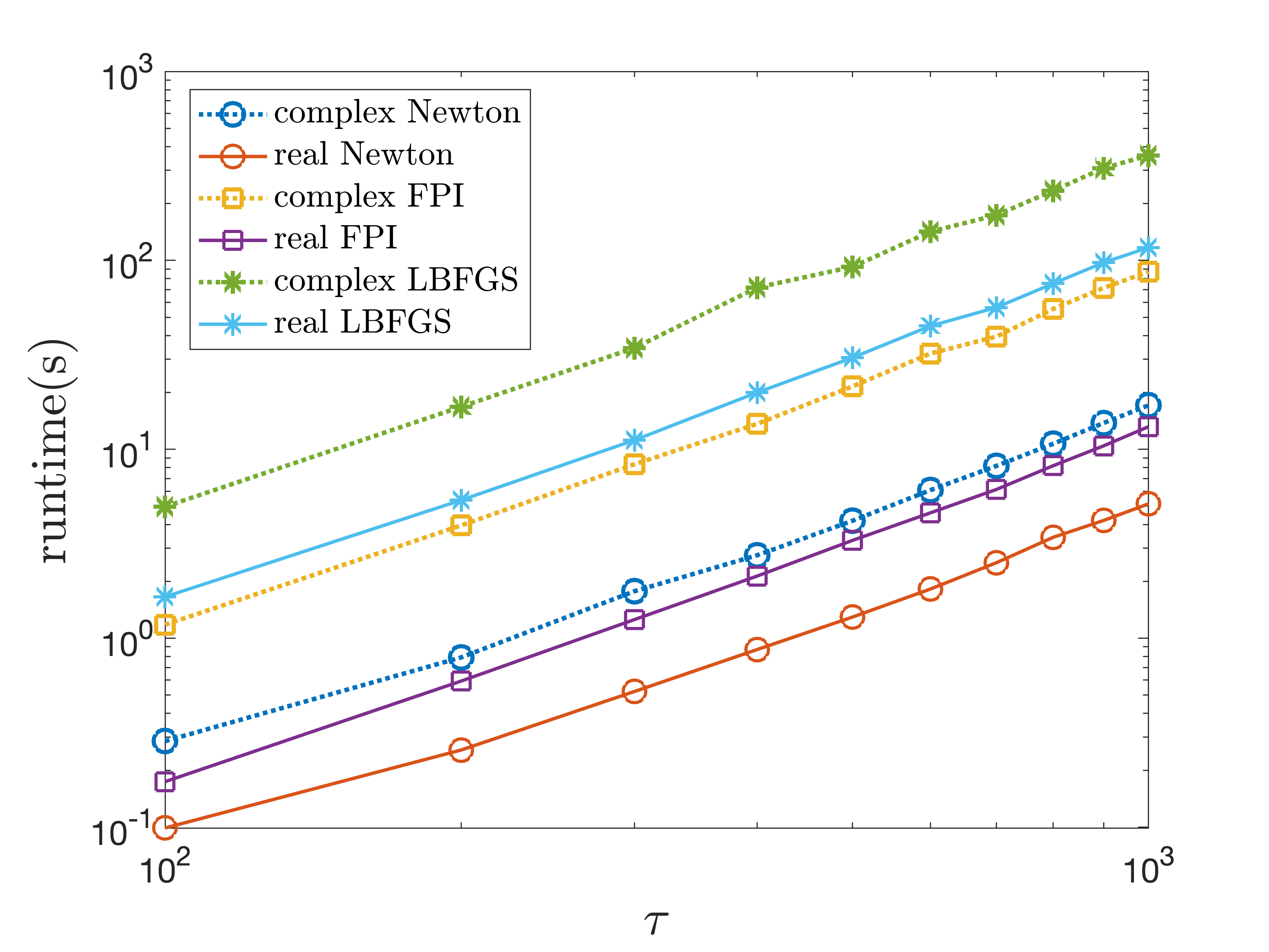}
        \caption{Comparing the runtime of iterative methods for finding phase factors using the real-number and complex-number arithmetic. The problem is set to quantum Hamiltonian simulation with variable $\tau$ parameters. The target polynomial is derived by truncating the Jacobi-Anger series with truncation error $\epsilon_0 = 1 \times 10^{-14}$. The maximal value of the target polynomial is scaled by a constant so that $\norm{f}_\infty = 0.9$.}
        \label{fig:HS_real_vs_complex}
\end{figure}

\begin{algorithm}[htbp]
\caption{computing the Jacobian matrix using the MPS/TT structure and the real-number arithmetic representation.}
\label{alg:DF_real}
\begin{algorithmic}
\STATE{\textbf{Input:} A set of reduced phase factors $\Phi$, its length $\wt{d}$ and its parity $p \in \{0, 1\}$.}
\STATE{Set $ d= 2\wt{d}-2  + p$ and initialize $\vg$ as a zero matrix of size $\wt{d}\times(2d+1)$.}
\FOR{$j = 0,\cdots, d$}
\STATE{Set $x_j=\cos\left(\frac{2\pi j}{2d+1}\right)$.}
\STATE{$\mc{N}_\mathrm{left}(x_j) =\mc{H}\prod_{i=\wt{d}-2}^{1}  \left(\mc{X}(x_j) \mc{Z}(\phi_i) \right)\mc{X}(x_j)$.}
\IF{ parity is even $(p = 0)$}
\STATE{$\mc{N}_\mathrm{right}(x_j) = (1,1,0)^\top$.}
\ELSE
\STATE{$\mc{N}_\mathrm{right}(x_j) = (x,x,\sqrt{1-x^2})^\top$.}
\ENDIF
\STATE{$g_{0,j}\leftarrow 2\mc{N}_\mathrm{left}\mc{Z}(\phi_0+\pi/4) \mc{N}_\mathrm{right}(x_j) $.}
\FOR{$ i= 1,\cdots, \wt{d}-1 $}
\STATE{$\mc{N}_\mathrm{left}(x_j)\leftarrow \mc{N}_\mathrm{left}(x_j) \mc{X}^{-1} (x_j) \mc{Z}(- \phi_i)$.}
\STATE{$\mc{N}_\mathrm{right}(x_j)\leftarrow  \mc{X}(x_j) \mc{Z}(\phi_{i-1})\mc{N}_\mathrm{right}(x_j)$.}
\STATE{$g_{i,j}\leftarrow 2 \mc{N}_\mathrm{left}(x_j)\mc{Z}(\phi_i+\pi/4) \mc{N}_\mathrm{right}(x_j) $.}
\ENDFOR
\ENDFOR
\STATE{Set $\vg_{i,j} \leftarrow \vg_{i ,2d+1-j}$, $j=d+1,\cdots, 2d$.}
\STATE{Compute $\vv_{il}\leftarrow \Re\left(\sum_{j=0}^{2d-1} \vg_{i,j} e^{-\I \frac{2\pi}{2d+1}l j}\right),l=0,\ldots,d$ by using FFT.}
\IF{parity is even $(p = 0)$}
\STATE{$\frac{\partial F(\Phi)}{\partial \phi_i} \leftarrow \frac{2}{2d+1}(\frac{\vv_{i0}}{2}, \vv_{i2}, \vv_{i4},\cdots, \vv_{id})$.}
\ELSE
\STATE{$\frac{\partial F(\Phi)}{\partial \phi_i} \leftarrow \frac{2}{2d+1}(\vv_{i1}, \vv_{i3}, \vv_{i5},\cdots, \vv_{id})$.}
\ENDIF
\STATE{\textbf{Output:} $D F(\Phi)$.}
\end{algorithmic}
\end{algorithm}
 In \cref{fig:HS_real_vs_complex}, we numerically demonstrate that using the real-number arithmetic formalism of QSP improves the time complexity of iterative methods by a constant prefactor. Notably, this improvement is not limited to Newton's method but also applies to other iterative methods for finding phase factors.

\end{document}